\documentclass[superscriptadress,citeauthorscript,aip,preprintnumbers,amsmath,amssymb,reprint,graphicx,amsmath,amssymb,twocolumn]{revtex4-2}
\usepackage{graphicx}
\usepackage{dcolumn}
\usepackage{bm}
\usepackage{color}
\usepackage{gensymb}
\usepackage[utf8]{inputenc}
\usepackage[T1]{fontenc}
\usepackage{mathptmx}
\usepackage{etoolbox}
\usepackage{url}
\usepackage{hyperref}
\usepackage[caption=false]{subfig}

\begin{document}


\title{A practical approach to calculating magnetic Johnson noise for precision measurements}
\author{N.~S.~Phan}
\email{nphan@lanl.gov}
\affiliation{Los Alamos National Laboratory, Los Alamos, New Mexico 87545, USA}

\author{S.~M.~Clayton}
\affiliation{Los Alamos National Laboratory, Los Alamos, New Mexico 87545, USA}

\author{Y. J. Kim}
\affiliation{Los Alamos National Laboratory, Los Alamos, New Mexico 87545, USA}

\author{T.~M.~Ito}
\email{ito@lanl.gov}
\affiliation{Los Alamos National Laboratory, Los Alamos, New Mexico 87545, USA}

\date{\today}

\begin{abstract}
 Magnetic Johnson noise is an important consideration for many applications involving precision magnetometry, and its significance will only increase in the future with improvements in measurement sensitivity. The fluctuation-dissipation theorem can be utilized to derive analytic expressions for magnetic Johnson noise in certain situations. But when used in conjunction with finite element analysis tools, the combined approach is particularly powerful as it provides a practical means to calculate the magnetic Johnson noise arising from conductors of arbitrary geometry and permeability. In this paper, we demonstrate this method to be one of the most comprehensive approaches presently available to calculate thermal magnetic noise.  In particular, its applicability is shown to not be limited to cases where the noise is evaluated at a point in space but also can be expanded to include cases where the magnetic field detector has a more general shape, such as a finite size loop, a gradiometer, or a detector that consists of a polarized atomic species trapped in a volume.  Furthermore, some physics insights gained through studies made using this method are discussed.
\end{abstract}

\maketitle

\section{\label{sec:introduction} Introduction}
Thermal current fluctuations in conducting materials caused by the voltage Johnson noise\cite{Johnson1928,Nyquist1928} generate fluctuating magnetic fields around the conductors. This magnetic Johnson noise\footnote{Note that the magnetic noise near the conductor surface that we are discussing in this paper differs from what is given by the blackbody radiation law.\cite{Henkel2005}} is becoming increasingly important in many precision measurements and applications because it can be a major source of interference, especially as measurement sensitivity improves. Magnetic Johnson noise was first studied in the context of biomagnetic measurements based on sensitive superconducting quantum interference device (SQUID) sensors \cite{Varpula1984,Nenonen1996} and for applications involving trapped atoms.\cite{Henkel2005} More recently, its impact on experimental searches for the permanent electric dipole moment (EDM) of atoms and the neutron has also been considered.\cite{Lamoreaux1999,Munger2005,Ayres2021} In these high precision experiments, magnetic Johnson noise can serve as a hindrance in multiple ways. When the experiment uses SQUID-based magnetometers,\cite{Ahmed2019} the sensitivity of these devices can be impacted by the magnetic Johnson noise arising from nearby conductive materials such as the electrodes and the materials in the Dewar or the magnetically shielded room.  In certain experiments, the noise from the conductors can even directly affect the spin precession frequency of the species of interest or that of the comagnetometer.\cite{Lamoreaux1999,Rabey2016,Ayres2021} 
 
Calculations of magnetic Johnson noise are typically performed using one of two methods: the direct method or reciprocal method.  In the direct method, the fluctuating current density is given by the conductivity of the material through the Johnson-Nyquist theorem\cite{Johnson1928,Nyquist1928} and the resulting fluctuating magnetic field is calculated either by solving Maxwell's equations\cite{Varpula1984,Nenonen1996} or by applying the Biot-Savart law.\cite{Lamoreaux1999,Sandin2010,Ready2021} The reciprocal method, on the other hand, makes use of the fluctuation-dissipation (F-D) theorem,\cite{Callen1951, Kubo1966} of which the Johnson-Nyquist theorem is a special case.  With this method, the magnetic Johnson noise (fluctuation) is determined from the power dissipated in the conductive body (dissipation) when it is driven by an oscillating magnetic field. 

Both methods have been used to derive analytic formulas or general expressions for some specific geometries.\cite{Varpula1984,Nenonen1996, Kasai1993, Roth1998, Clem1987, Sidles2003, Lee2008} However, calculating the magnetic Johnson noise using the direct method for a general case is a daunting task [see Eqs. (17) and (19) of Ref.~\onlinecite{Varpula1984}]. When appropriate, one can resort to the static approximation,\cite{Lamoreaux1999} that is to limit the solution to zero frequency, and employ numerical methods\cite{Ready2021,Ayres2021} to obtain a solution. On the other hand, with the F-D theorem based method, the magnetic Johnson noise caused by the conductive body with an arbitrary shape can be calculated for various frequencies in a more straightforward manner by taking advantage of the recent advancement in finite element tools, such as COMSOL,\cite{COMSOL} as pointed out by e.g. Refs.~\onlinecite{Lee2008} and \onlinecite{Iivanainen2021}. 

However, in all of the applications of the F-D theorem based method that the authors have seen, the evaluation of the magnetic Johnson noise is performed at an observation point in space, using an infinitesimally small current loop. The purpose of this paper is to show, using some concrete examples, that the combined method of the F-D theorem and finite element tools has greater, more general applicability. That is, the method is also applicable to cases where the magnetic field detector (e.g., a pickup loop) i) has a finite area with an arbitrary shape, ii) is arranged as a gradiometer, and iii) has a finite volume (e.g., polarized atomic species trapped in a glass bulb).  In addition, we will discuss additional physics insights gained through studies made using this method. 

\section{\label{sec:FDtheorem} Method}

In the reciprocal method, when evaluating the magnetic Johnson noise at a given observation point near the conductor, a small hypothetical current loop with an oscillating current $I(t)=I_0 \sin{\omega t}$ and area $A$ is placed at that location.  The resulting oscillating magnetic field incurs a time-averaged power dissipation $P$ in the conductor due to eddy currents. The generalized Nyquist relation,\cite{Callen1951} which is an expression of the F-D theorem,\cite{Callen1951, Kubo1966, Sidles2003, Lee2008} can be used to show that the power dissipation $P$ and the magnetic Johnson noise $B_{n}$ are related by
\begin{equation}
\label{eq:F-D}
B_{n}(f) =\frac{
\sqrt{4kT}\sqrt{2P(f)}
}
{2\pi fANI}.
\end{equation}
Here, $T$ is the temperature of the conductor, $k$ is the Boltzmann constant, $N$ is the number of turns in the current loop, and $\omega = 2 \pi f$ is the angular frequency of the driving current. The magnetic dipole moment of the current loop is $\vec{p} = NI\vec{A}$. When a finite sized magnetic field detector is used, the current loop is replaced by a wire loop of the appropriate geometry, such as a coil of finite area, multiple current loops arranged as a gradiometer, or a coil geometry that encloses a finite volume. The magnetic dipole moment of the corresponding geometry then replaces that for the simple current loop in Eq.~\ref{eq:F-D}.

When calculating the magnetic Johnson noise employing the F-D theorem based method together with a finite element analysis tool, the task reduces to the following procedure:
\begin{enumerate}
    \item Set up a relevant geometry in the finite element analysis tool.
    \item Place a wire loop of an appropriate geometry at the location where the noise is to be evaluated.
    \item Generate an oscillating current in the wire loop and calculate the power dissipation in the conductor(s).
    \item Use Eq.~(\ref{eq:F-D}) to calculate the magnetic Johnson noise from the calculated power dissipation.
\end{enumerate}
In Step 2, for a point evaluation of the magnetic Johnson noise, the wire loop needs to be small; i.e., the diameter of the loop is smaller than the distance between it and the conductor surface(s). On the other hand, for evaluation of the magnetic Johnson noise averaged over a finite sized loop, the wire loop needs to have the same geometry as the loop used in the measurement. For a first-order gradiometer, two loops with currents flowing in the opposite direction need to be used. For a coil with a finite length, the loop is replaced by such a coil (e.g., a solenoid of the given length). If a volume average needs to be calculated, for example, to study the effect of magnetic Johnson noise on the performance of a magnetometer made of trapped polarized atoms, a coil geometry that generates a uniform field distribution within the volume is required. In such a geometry, an oscillating magnetic field at each location inside the coil produces an oscillating current with an equal amplitude in the coil wire.

This method, henceforth, referred to as the "F-D + FEM" method, gives the correct frequency dependence automatically.  For all of the results presented in this paper, COMSOL Multiphysics 6.2\cite{COMSOL} was used as the FEM software tool.  But because the method does not depend on any features exclusive to COMSOL, all of the results presented are general, and are applicable to any FEM tools, commercially available or otherwise.  The advantage of using a commercially available tool is that the task of setting up the geometry and performing the power loss computation is made considerably straightforward, in that the level of effort is no more laborious than typical FEM-based simulations.  But it is important to note that the applicability of the results to a particular experiment depends on the accuracy of the inputs into the simulation.  Rather than using nominal literature values or default ones from the material database found in the FEM software package, material properties should be determined experimentally when appropriate and feasible.

\begin{figure}
    \centering
    \includegraphics[width=0.48\textwidth] {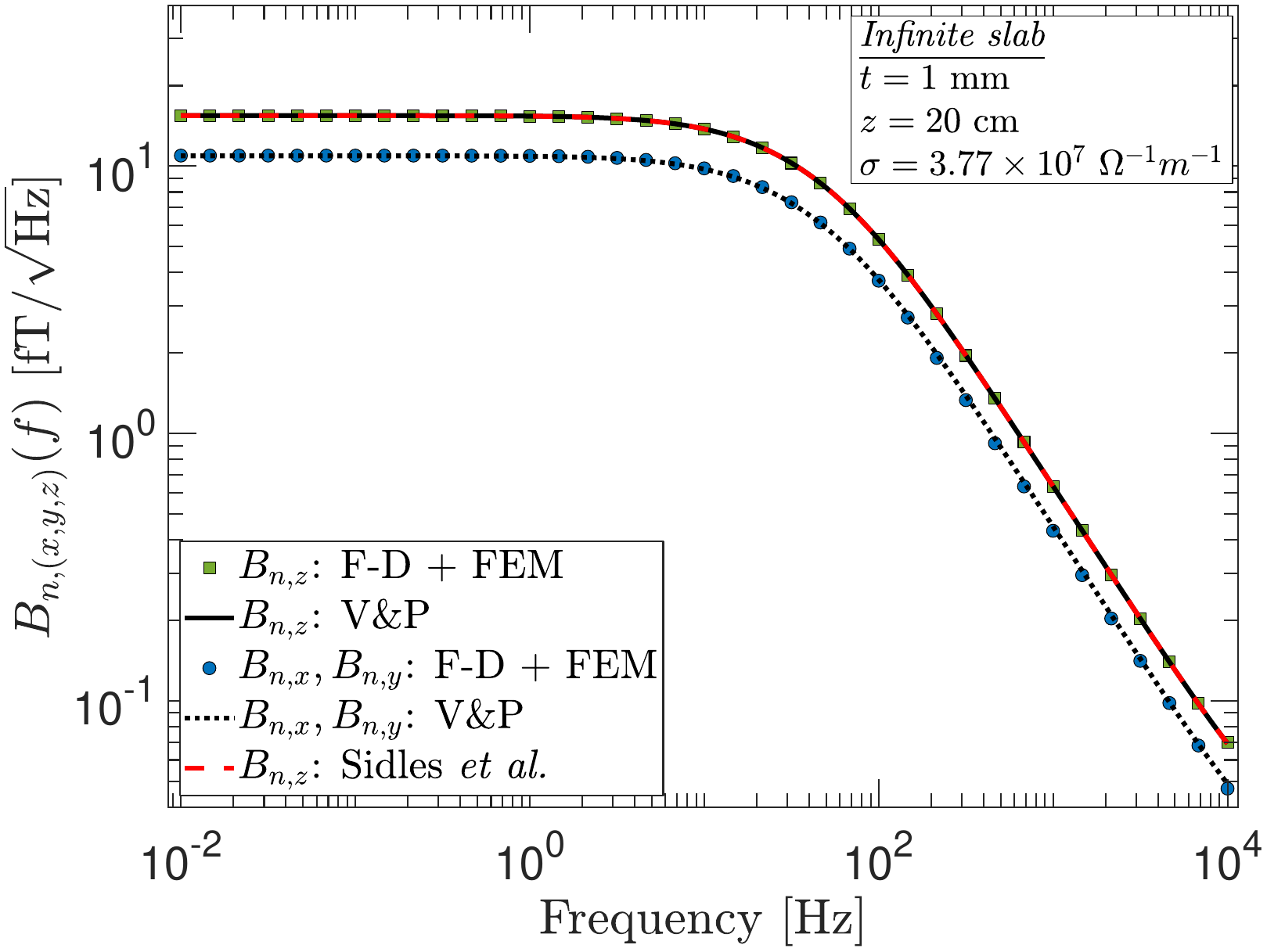}
    \caption{A comparison of the magnetometer noise components, $B_{n,(x,y,z)}(f)$, numerically computed from the expressions derived by Varpula and Poutanen (V\&P)\cite{Varpula1984}  and Sidles \textit{et al.}\cite{Sidles2003} with those obtained from the F-D theorem and finite element method (F-D + FEM) for the case of an infinite slab geometry.}
    \label{fig:infinite-slab-magnetomer}
\end{figure}

\section{\label{sec:comparison} Validation of the method for various situations}
\subsection{Simple conductor geometry: nonmagnetic}\label{sec:infinite-slab-non-magnetic}
We start by validating the results of the F-D + FEM method against those obtained from numerical calculations for simple conductor geometries. In Fig.~\ref{fig:infinite-slab-magnetomer} we show a comparison of the results obtained from numerical integration of the expressions from Varpula and Poutanen\cite{Varpula1984} and Sidles \textit{et al.}\cite{Sidles2003} for the case of an infinite slab against those obtained by the F-D + FEM method. Plotted are the components of the field noise as a function of the frequency $f$ at the given position $z$.  Here and throughout, unless specified otherwise we take the conductor slab to lie in the $x\text{--}y$ plane with $\hat{z}$ normal to its surface.  The conductor is taken to be aluminum with temperature $T=293$~K, thickness $t=1$~mm, conductivity $\sigma = 3.77 \times 10^{7}$ $\Omega^{-1} {\rm m}^{-1}$, permeability equal to that of vacuum $\mu_0$, and the distance $z$ of 20~cm is measured from its top surface (i.e., the closest distance between the conductor and the location where the noise is evaluated).  As shown in Fig.~\ref{fig:infinite-slab-magnetomer}, there is excellent agreement between the results of the F-D + FEM method and the numerically computed results of Varpula and Poutanen\cite{Varpula1984} and Sidles \textit{et al.}.\cite{Sidles2003}  Importantly, a numerical survey by Sidles \textit{et al.} shows the latter two methods yield identical predictions in the same physical regime even though they are derived from different physical models (direct vs. reciprocal).\cite{Sidles2003}  This implies that a comparison of the F-D + FEM method with either of those methods should suffice.  Furthermore, this exercise demonstrates agreement for the different components of the field noise and confirms the relationship amongst field components from Varpula and Poutanen\cite{Varpula1984} that
$$B_{n,x}(z, f) = B_{n,y}(z, f) = \frac{1}{\sqrt{2}}B_{n,z}(z, f).$$
Of particular note is that an alternative but incorrect relationship of the form $B_{n,x} = B_{n,y} = \sqrt{3/2}B_{n,z}$ is obtained by some authors when calculating the magnetic noise using the Biot-Savart law for the infinite slab geometry.  As suggested by Henkel,\cite{Henkel2005, Henkel1999} this is due to a difficulty in handling the boundary condition at the surface of the conductor when modeling the noise source as stationary currents in vacuum.

\begin{figure}
    \centering
    \includegraphics[width=0.48\textwidth] {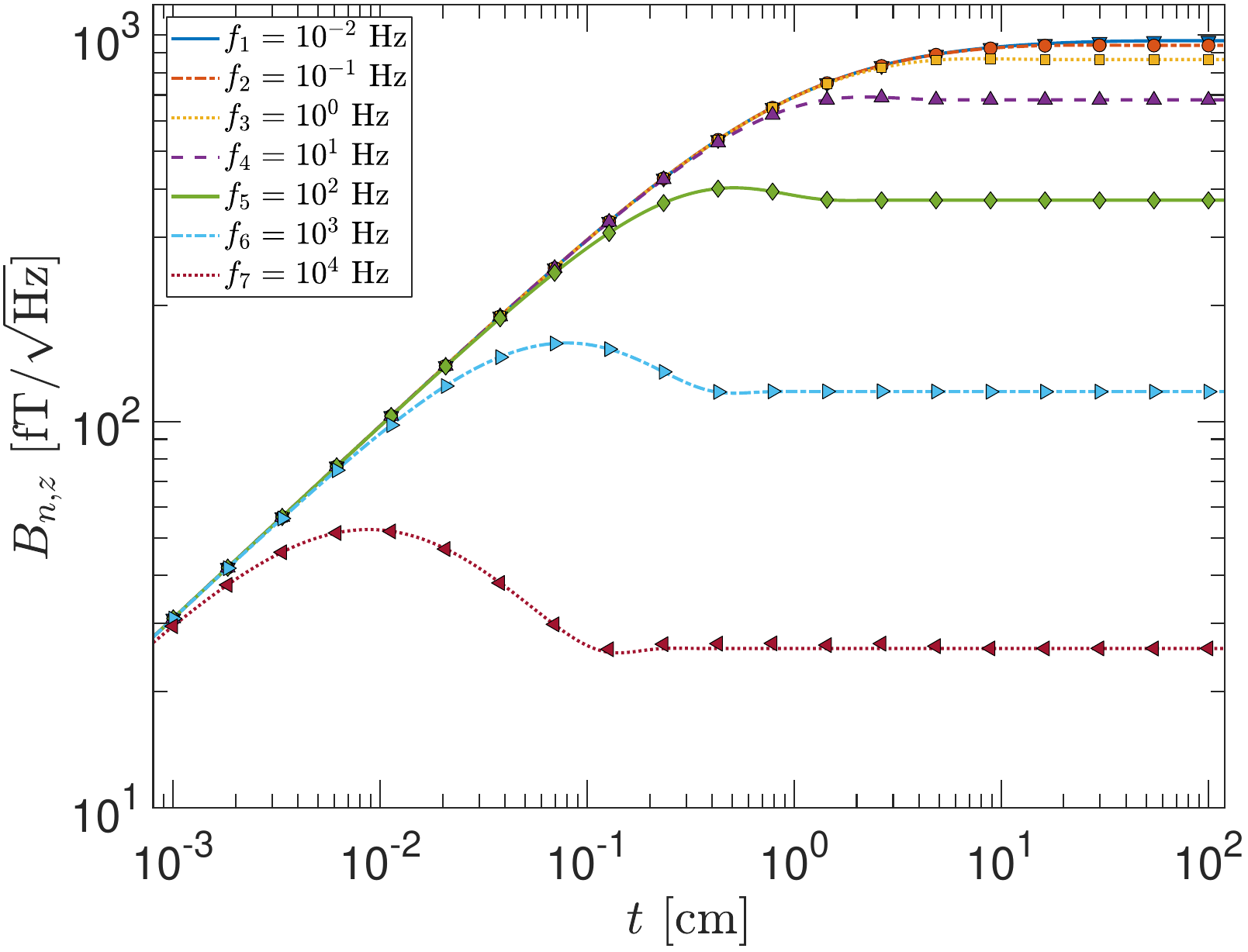}
    \caption{The magnetometer noise $B_{n,z}$ as a function of $t$, the thickness of the infinite slab, at a fixed point of observation $z =1$~cm for several frequencies. The curves show the numerical results from Varpula and Poutanen\cite{Varpula1984} while the markers are results obtained with the F-D + FEM method.}
    \label{fig:slab-thickness-frequency}
\end{figure}

Some investigators adopt the approach of using the Biot-Savart law to estimate the noise because it is not so straightforward to implement the formulation of Varpula and Poutanen or that of Sidles \textit{et al.} for an arbitrary geometry.  This is further motivated by the recognition that in most experiments the skin depth, $\lambda = 1/\sqrt{\pi \mu_{0} \sigma f}$, is the largest length scale when the frequency of interest is small.  In this so-called quasi-static regime, one can make a simplifying assumption that the eddy currents in the conductor bulk can be neglected.  Although useful, such an approximation limits the range of applicability and accuracy of the calculation.  The F-D + FEM method, however, does not suffer from these limitations.  In Fig.~\ref{fig:slab-thickness-frequency}, we show that the method produces the expected results across the different regimes encountered in experiments: 
\begin{itemize}
\item $\lambda \gg  t$
\item $\lambda \sim t$
\item $t \gg \lambda,z$
\end{itemize}
where $\lambda$ is the skin depth, $t$ the thickness of the conductor, and $z$ the distance between the conductor and the observation point.

The small deviations of 1--2\% between the curves and some of the points are due to a combination of the mesh granularity used to model the geometry in the FEM software and the truncation that must be made to reduce the infinite slab to a finite body for computation.  The latter is never an issue in practice when calculating the magnetic Johnson noise for an experimental apparatus because no components has an infinite size.  In principle, the level of precision and accuracy can be made much higher but at the expense of computational resources and time.  The FEM simulations presented in this paper used a standard workstation computer (16 CPU cores, 128 GB of memory) and the computation time for each simulation varied between 1-2 hour to a few minutes.  But in practice, uncertainties in material properties and geometry may be the primary limitations on the accuracy of the results rather than computational factors.  Furthermore, the level of precision that is of relevance to the experimental work is also an important consideration in deciding on a reasonable amount of computational resources and time to devote to the simulation.

Immediately apparent from Fig.~\ref{fig:slab-thickness-frequency} are three distinct regions.  In the limit that $t$ is much less than the skin depth at the highest frequency of interest, the noise is approximately independent of the frequency.  This is the regime in which the static approximation is applicable.  On the opposite extreme is the region in which $t$ becomes the largest length scale and is much greater than both $\lambda$ and $z$.  We see in this regime that the noise approaches a constant value and does not change beyond that point.  This is consistent with the intuitive view that the field from the lower layers of the conductor is effectively screened by the top layers.  In the intermediate regime where $t$ is comparable to $\lambda$, the noise is attenuated and reaches a maximum around $\lambda \sim \sqrt{zt}$.

\subsection{\label{sec:high-permeability}Simple conductor geometry: high-permeability}
As discussed by Lee \& Romalis in Ref.\onlinecite{Lee2008}, the F-D theorem based method can be used for the calculation of noise arising from multiple physical origins. Therefore, it is straightforward to extend the method to noise calculations for magnetic and high-permeability materials as long as the material properties and power loss mechanisms are accurately modeled in the FEM software.  For high-permeability metals, the two primary sources of power-loss and, hence, magnetic noise are eddy current heating due to the conductivity of the material and hysteresis loss associated with magnetic domain fluctuations.  The former depends on the conduction of the material whereas the latter depends on $\mu^{'}$ and $\mu^{''}$, which are the real and imaginary parts of the complex permeability $\mu = \mu^{'} - i\mu^{''}$.  If the imaginary part can be neglected, then $\mu = \mu_r \mu_0$, where $\mu_r$ is called the relative permeability.

\begin{figure}
    \centering
    \includegraphics[width=0.48\textwidth] 
{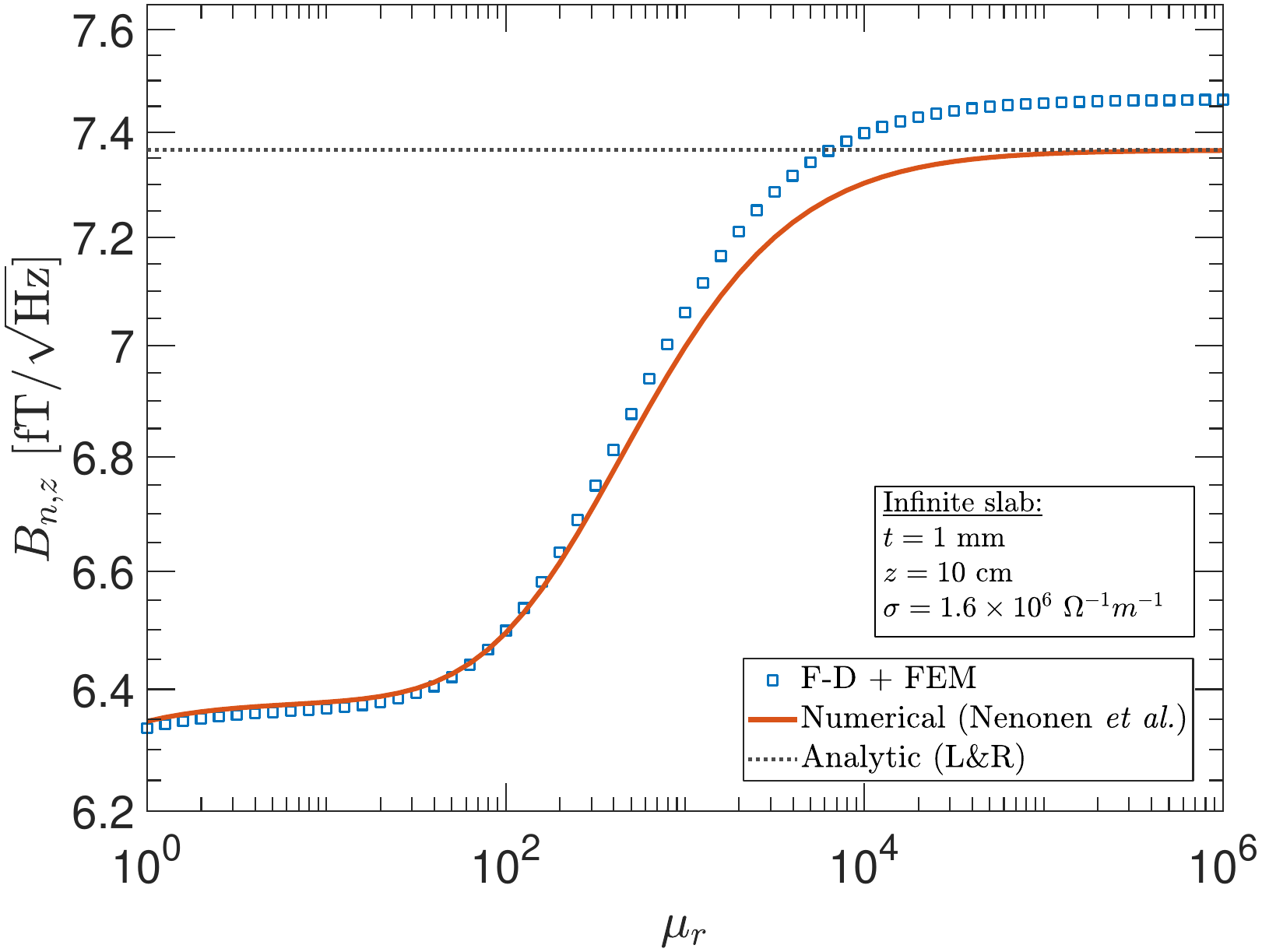}
    \caption{The magnetometer noise as a function of the relative permeability $\mu_r$ at a fixed observation point of $z =10$~cm above a high-permeability infinite slab. Shown are the results in the zero frequency limit obtained with the F-D + FEM method (blue-squares), the numerical evaluation of the formulas from Nenonen \textit{et al.}~\cite{Nenonen1996} (solid-red line), and the analytic result from Lee \& Romalis (L\&R)~\cite{Lee2008}(dotted-black line) that is valid in the high-permeability limit ($\mu_r \gg 1$ and $\mu_r t/z \gg 1$).}
    \label{fig:infinite-slab-high-perm-noise-vs-mur}
\end{figure}

In Fig.~\ref{fig:infinite-slab-high-perm-noise-vs-mur}, we show calculations of the magnetic noise in the zero frequency limit due to resistive losses (eddy current heating) in a $t = 1$-mm thick high-permeability infinite slab as a function of $\mu_r$.  The observation point is located at a distance of $z = 10$~cm and the conductivity is taken to be $\sigma = 1.6 \times 10^{6}$~$\Omega^{-1}~{\rm m}^{-1}$. The results in Fig.~\ref{fig:infinite-slab-high-perm-noise-vs-mur} show agreement at the percent-level or better between the numerical and F-D + FEM calculations over the entire range of permeabilities.  Further, the analytic result of Lee \& Romalis, which is valid when the permeability satisfies the conditions: $\mu_r \gg 1$ and $\mu_r t/z \gg 1$, also shows good agreement with the other two methods in the high-$\mu_r$ range.  In the FEM modeling, a simple constitutive relation with $\vec{B} = \mu_o \mu_r \vec{H}$ is assumed. However, in practice, many materials can exhibit nonlinear behavior and their modeling is more complicated.  The simplifying assumption adopted here is only valid for certain materials and within certain regimes but is useful for the purpose of demonstrating the F-D + FEM method.

\begin{figure}
    \centering
    \includegraphics[width=0.48\textwidth] 
{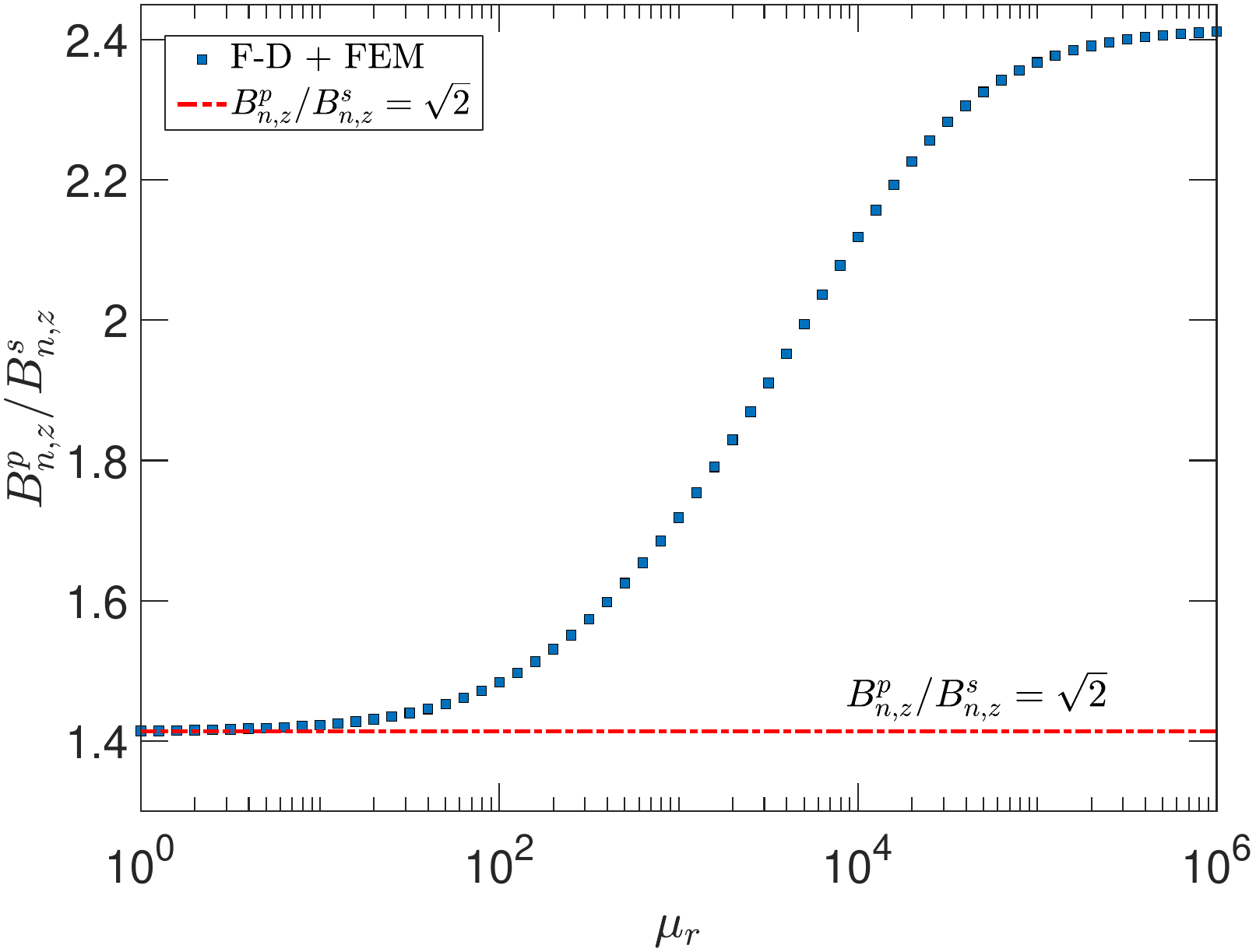}
    \caption{The ratio of the noise $B_{n,z}^{p}$ for an observation point in the middle of two parallel infinite slabs to the noise $B_{n,z}^{s}$ at the same point with one slab present as a function of $\mu_r$. The conductivity and thickness of the slab are taken to be $\sigma = 1.6 \times 10^{6}$~$\Omega^{-1} {\rm m}^{-1}$ and $t = 1$~mm, respectively, and the distance is $z = 10$~cm.  All calculations are done for the zero frequency limit.}
    \label{fig:ratio_quadrature_sum}
\end{figure}

An important general result for high-permeability material is that the total noise from an object made of such material cannot be obtained by a quadrature summation of the noise from the individual components comprising it.\cite{Lee2008}  Lee \& Romalis\cite{Lee2008} assert that this is due to the effect of image currents.  Furthermore, when a high-permeability material is used in conjunction with a conductor, the reflection effect must also be considered.\cite{Maniewski1985,Nenonen1996}  We demonstrate these features with the F-D + FEM method by considering a geometry consisting of two parallel infinite slabs with the observation point located in the middle of the gap between the slabs.  

Figure~\ref{fig:ratio_quadrature_sum} shows the ratio of the noise with two slabs to the noise when only one slab is present as a function of the relative permeability.  If the total noise for the two slab configuration obeys the quadrature summation rule, the expected value would be $\sqrt{2}$ larger than the case with only one slab.  However, the calculations show that this is true only for low permeabilities below $\sim$100.  In the high permeability limit, the results obtained with the F-D + FEM method are incompatible with the conventional, yet incorrect, assumption of $\sqrt{2}$.  They are, however, consistent with the analytic results of Lee \& Romalis\cite{Lee2008} which give a ratio of 2.54 for a closed end cylinder of low aspect ratio ($H/2R = 0.05$), where $H$ is the height and $R$ the radius, so the noise is dominated by the end caps of the cylinder.  Moreover, the effect is also present, but at a different level, if one of the high-permeability slabs is replaced by a conductor with $\mu_r = 1$. Therefore, a more cautious approach to noise subtraction or addition for geometries containing high-permeability materials is most appropriate.

\subsection{\label{sec:freq-depend}Frequency dependence}

\begin{figure}
    \centering
    \includegraphics[width=0.48\textwidth] 
{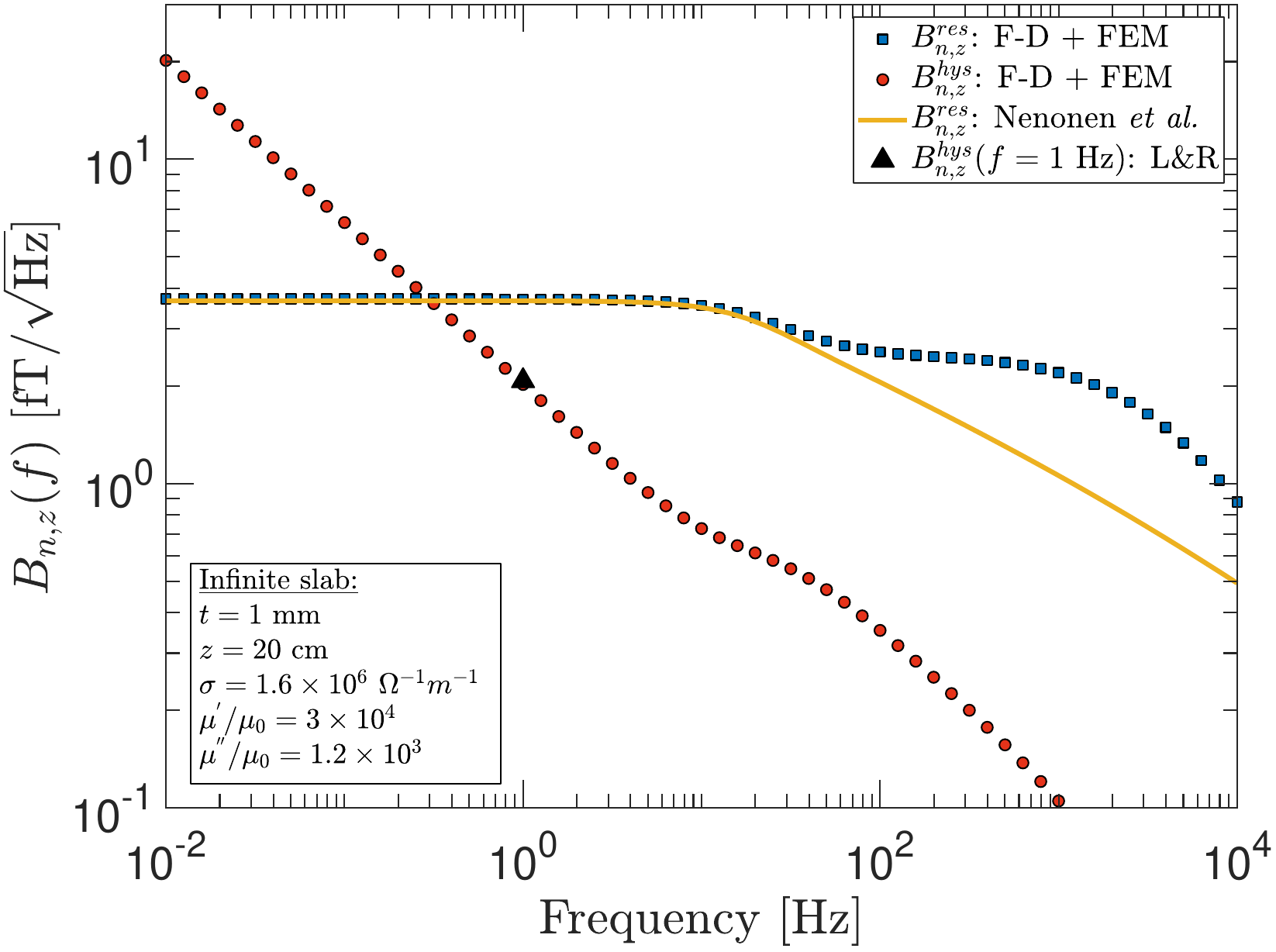}
    \caption{The magnetometer noise as a function of frequency at a fixed observation point of $z =20$~cm above a high-permeability infinite slab. Shown are the noise components due to conductivity of the conductor ($B_{n,z}^{res}$) and hysteresis ($B_{n,z}^{hys}$) as calculated using the F-D + FEM method.  The numerical and analytic results of Nenonen \textit{et al.}\cite{Nenonen1996} and Lee \& Romalis (L\&R)\cite{Lee2008} are also shown, respectively.}
    \label{fig:infinite-slab-high-perm-noise-vs-freq}
\end{figure}

In general, materials of different permeabilities exhibit rather distinct frequency dependence.  For \textit{thin} ($z \gg t$) nonmagnetic infinite slabs such as the one considered in Sec.~\ref{sec:infinite-slab-non-magnetic}, three regions are present and the general frequency dependence is given by:
$B_{n}^{}(f): f^{0} \rightarrow f^{-1} \rightarrow f^{-3/4}$, where the transition from the $f^{0}$ region to the $f^{-1}$ region occurs around the roll-off frequency $f_{c} = 1/4\mu_{0}\sigma z t$ due to inductive screening effects.  But at frequencies much higher than $1/2\mu_{0}\sigma t^2$, the skin depth becomes important and the noise scales as $\propto f^{-3/4}$. For the example shown in Fig.~\ref{fig:infinite-slab-magnetomer}, the initial roll-off and skin-depth frequencies are $\approx 26.4$~Hz and $\approx 1.05\times 10^{4}$~Hz, respectively.  This implies that the skin-depth effect may not necessarily be most pertinent in all situations, yet this is often assumed when adopting the static approximation.

With high-permeability slabs, the frequency dependence is rather more complicated.  In Fig.~\ref{fig:infinite-slab-high-perm-noise-vs-freq}, we show F-D + FEM calculations for the frequency dependence of the magnetic noise due to a high-permeability infinite slab possessing the same conductor and magnetometer geometry as that used in Sec.~\ref{sec:infinite-slab-non-magnetic}.  The contributions to the noise from the resistive (eddy current heating) and hysteresis loss components are shown (blue and red markers).  The solid-curve and black-triangle show the numerical and analytic results from Nenonen \textit{et al.}\cite{Nenonen1996} and Lee \& Romalis,\cite{Lee2008} respectively.  It can be seen that in the low frequency regime, the noise resulting from hysteresis losses dominates and exhibits a frequency dependence of $f^{-1/2}$, in agreement with hysteresis power loss being proportional to the frequency.  The resistive component on the other hand is the primary noise source for $f > f_{\rm{mag}} = 3\tan{\delta} / 2 \pi \mu_r \mu_0 \sigma t^2 $, where $\tan \delta = \mu^{''}/\mu^{'}$ is the ratio of the imaginary to the real part of the complex permeability.  (See Eqs. (9) and (10) of Lee and Romalis\cite{Lee2008} and the text in between for further details.)  For the example considered in Fig.~\ref{fig:infinite-slab-high-perm-noise-vs-freq}, $f_{\rm{mag}} \approx 0.3$~Hz.  Hence, the frequency dependence for the total noise is given by: $B_{n}^{\textrm{tot}}(f): f^{-1/2} \rightarrow f^{0} \rightarrow f^{-\gamma} \rightarrow f^{-\xi}$,
where $\gamma \in [ 1/4, 1]$.  The parameter $\gamma$ is associated with the roll-off due to the skin-depth effect and increases with decreasing permeability.  

Interestingly, the numerical results are seen to deviate from those obtained with the F-D + FEM method for frequencies greater the initial roll-off, but the deviation only becomes substantial for $\mu_r > 10^3$ such as for the example in Fig.~\ref{fig:infinite-slab-high-perm-noise-vs-freq}.  For this reason, we leave the range for the parameter $\xi$ unspecified.  The disagreement is, however, suggestive of a power loss mechanism that is not accounted for in the numerical calculations.  For example, one such mechanism, referred to as excess loss, is the result of the large-scale motion of magnetic domain walls and is distinct from hysteresis loss.~\cite{Brailsford1948, Williams1950,Pry1958}  The precise reason for the disagreement remains unclear, nonetheless, and we leave this matter for future investigation.

\subsection{\label{sec:finite-loop}Correlation in a finite-size loop}

In many applications, spatial correlations are an important consideration because the magnetic noise is not simply measured at a single point.  Rather, coils of finite size are used and so the magnetic field lines will transverse the entire area enclosed by the coil(s), whether within a given coil or across multiple coils.  Thus, the noise is correlated within the area or volume enclosed by the coil(s).  We show that such correlations are appropriately accounted for by the F-D + FEM method through comparison with the results of Nenonen $\textit{et al.}$.\cite{Nenonen1996}  There, the authors compared the noise evaluated at a point with that for a finite size coil above an infinite conductor slab.  Generally, the magnetic noise field is defined as $B_{n,z}(\vec{r}) = \left(\overline{B_{z}(\vec{r})^{2}}/\Delta f \right)^{1/2}$, and so the noise for a finite detector, $B_{n,z}^{c}$, is obtained by integrating the field $B_{z}(\vec{r})$ over the enclosed area or volume of the detector.

\begin{figure}
    \centering
    \includegraphics[width=0.48\textwidth] {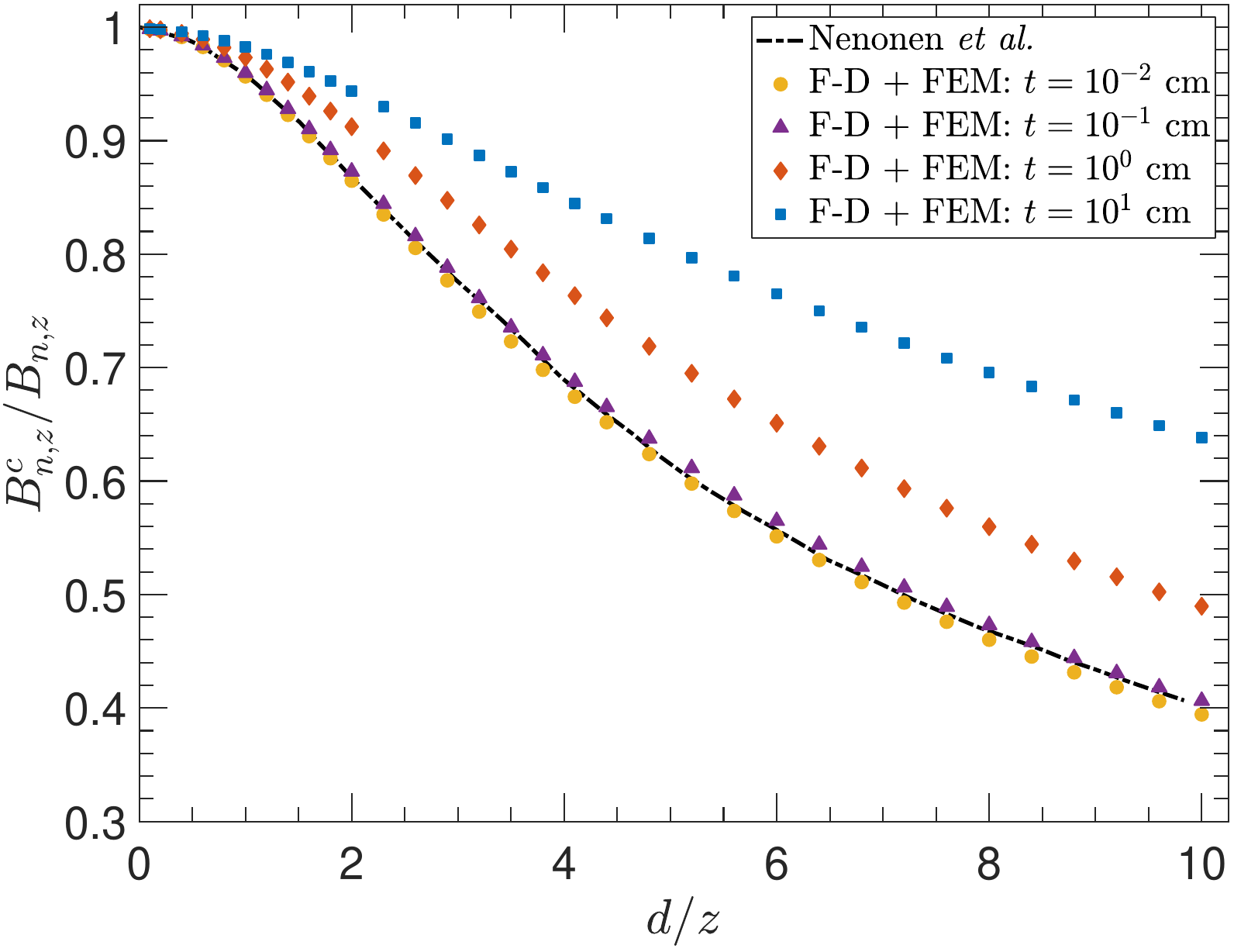}
    \caption{The ratio of the noise $B_{n,z}^{c}$ calculated for a finite-size circular coil of diameter $d$ and distance $z$ from the conductor surface to the noise $B_{n,z}$ calculated for a point-size coil at the same distance.  The black dot-dashed curve is from Nenonen \textit{et al}.\cite{Nenonen1996} while the points are calculated with the F-D + FEM method for several conductor thicknesses. The calculations take $\sigma = 3.77 \times 10^{7}$ $\Omega^{-1} {\rm m}^{-1}$ and $f=0.01$~Hz.}
    \label{fig:coil-size-distance}
\end{figure}

Figure \ref{fig:coil-size-distance} shows the ratio of the noise for a finite-size circular coil of diameter $d$ and distance $z$ placed parallel to the surface of a planar conductor to the noise measured by a point-size coil at the same distance for $f=0.01$~Hz.  The level of correlation in the noise at different $x\text{--}y$ locations with the same $z$ coordinate is contained within this ratio.  Points with larger separation in $x$ or $y$ (large $d/z$) have lower correlations, whereas nearby points are highly correlated.

The F-D + FEM method is shown to reproduce the numerical results of Nenonen $\textit{et al.}$\cite{Nenonen1996} in the same regime -- that of thin conductors ($t\leq 1$~mm).  Using the method, we also highlight the fact that the correlation can change for thicker conductors (i.e., the fields at two spatial locations with the same $z$ coordinate are more correlated with increasing conductor thickness).  Similar behavior is exhibited by the other field components, and in general, the correlations are also dependent on the frequency.

Furthermore, we note that the calculations in Fig.\ref{fig:coil-size-distance} take $f=0.01$~Hz and $\sigma = 3.77 \times 10^{7}$ $\Omega^{-1} {\rm m}^{-1}$, corresponding to a skin depth of $\sim82$~cm.  This is worth noting because it demonstrates that changes in the correlation can occur at length scales much smaller than the skin depth, so it is therefore prudent to consider all the appropriate length scales in the experimental arrangement, and not just the skin depth as is often done in practice.

\subsection{\label{sec:gradiometers}Gradiometer}
\begin{figure}
    \centering
    \includegraphics[width=0.48\textwidth] {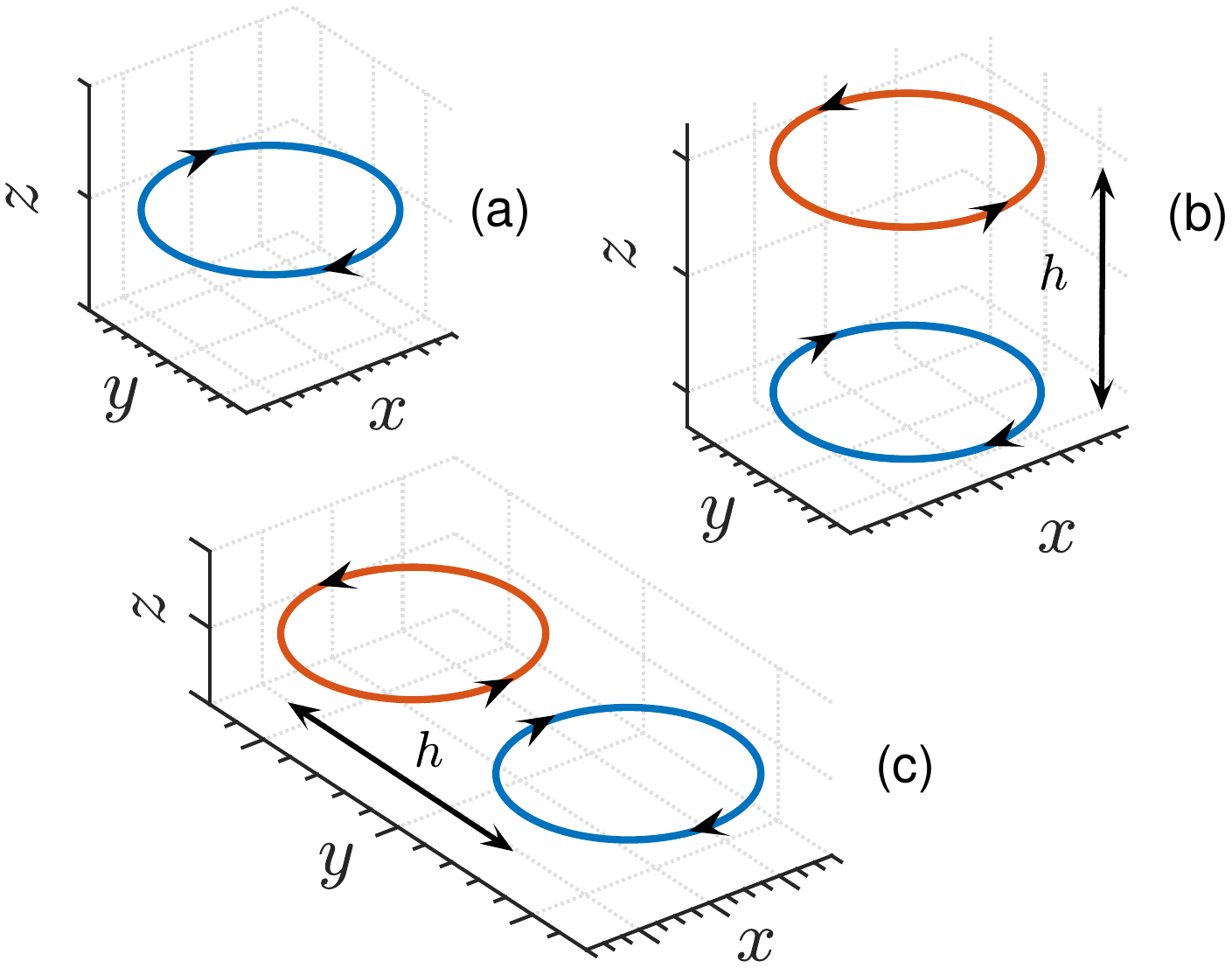}
    \caption{Schematic diagrams of magnetometer and gradiometer coil geometries: (a) magnetometer, (b) symmetric first-order axial gradiometer and (c) symmetric first-order planar gradiometer.  The other standard configurations are obtained by 90$\degree$ rotations about the $x$ and $y$ axes.  The arrows indicate the current loop senses, and $h$ is the gradiometer baseline.}
    \label{fig:gradiometer-geom}
\end{figure}

\begin{figure}
    \centering
    \includegraphics[width=0.48\textwidth] {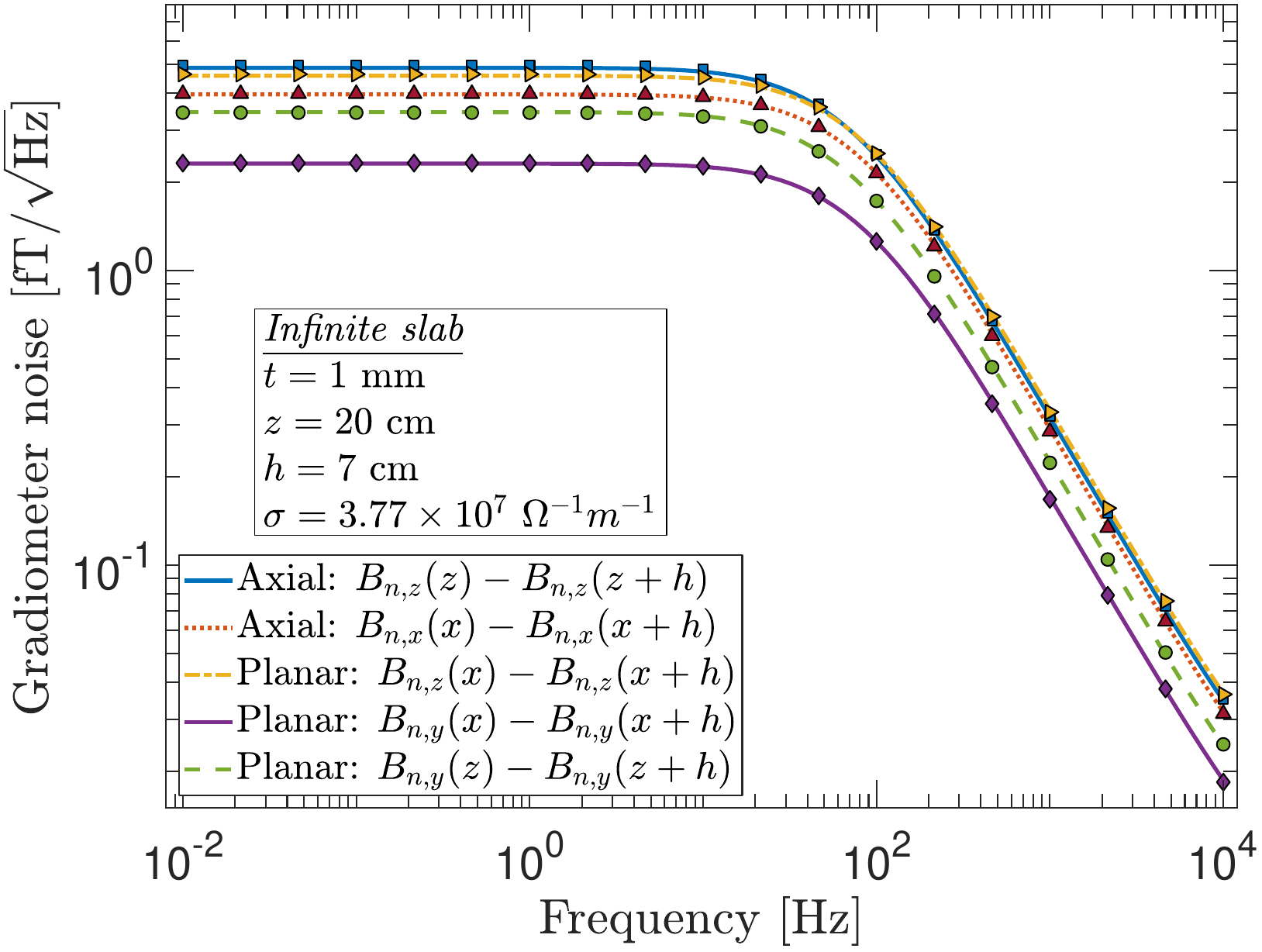}
    \caption{A comparison of the gradiometer noise, first-order axial and planar, obtained through numerical integration of the results from Varpula and Poutanen (V\&P)\cite{Varpula1984} and Nenonen \textit{et al.}\cite{Nenonen1996} (curves) with those calculated with the F-D + FEM method (markers) for the case of an infinite aluminum slab geometry. Here, $z$ and $z+h$ are the distances from the conductor surface to the near and far pickup loops, respectively, and $h$ is the separation between the two loops, also referred to as the baseline of the gradiometer.  Note that the plot shows the field noise but the equivalent field gradient noise can be found by dividing the curve by the baseline.}
    \label{fig:infinite-slab-gradiometer}
\end{figure}

Correlations also play an important role in differential measurements made by gradiometers.  A gradiometer consists of two or more pickup loops separated by some baseline, $h$ (refer to diagrams in Fig.\ref{fig:gradiometer-geom} for different gradiometer configurations).  In a two pickup loop gradiometer (also called first-order or first-derivative), a gradient component of the form $\partial B_{i} /\partial x_{j}$ is measured. The gradiometer is an axial type when $i = j$ and is planar when the indices represent orthogonal axes, but in general can be of neither type. \cite{Zimmerman1977} When a gradiometer is placed near a body of conductor, the two loops sense magnetic fields that are partially correlated. Using the direct method, Varpula and Poutanen\cite{Varpula1984} worked out expressions for the magnetic Johnson noise measured by a gradiometer for simple conductor geometries, and Nenonen \textit{et al.} \cite{Nenonen1996} extended these calculations to additional gradiometer configurations. Similar results are obtained through the reciprocal method by Clem.\cite{Clem1987}  The F-D + FEM method reproduces their results as seen in Fig.~\ref{fig:infinite-slab-gradiometer}.  Here, the loops are assumed to be identical, each having magnetic dipole moment $\vec{p}$, and the gradiometer noise is obtained from $\sqrt{4kT}\sqrt{2P}/(2\pi f p)$.\cite{Lee2008}  However, in the most general setup, it is more natural to consider the current noise in the pickup loops, $I_n$, rather than the magnetic field noise.  In which case, one can recast Eq.~\ref{eq:F-D} in terms of the gradiometer inductance $L_{G}$ when it is connected to a SQUID of input inductance $L_{i}$ as $I_n(f) = \sqrt{4 k T} \sqrt{2 P(f)}/(2 \pi f L_{0} I_{exc})$, where $P(f)$ is the power dissipated at frequency $f$ for $I_{exc}$ current in the pickup loop and $L_{0} = L_G + L_i$ is the total inductance.\cite{Clem1987}  The advantage of the F-D + FEM method is that it is straightforward to extend its applicability to very general coil geometries such as higher-order axial and planar gradiometers or for gradiometers of neither type, and for cases where the gradiometer is placed near a conductor of arbitrary size and shape. In general, correlated measurements involving multiple pickup loops, including those in a magnetometer array, can be modeled with the F-D + FEM method.

\subsection{Finite geometries: thin and thick conductors}

In practice, it is not always possible to approximate the experimental arrangement using some of the idealized geometries for which analytic formulas exist.  Up until this point, we have validated the F-D + FEM method against the numerical calculations of expressions applicable to the canonical infinite slab geometry.  We now turn our attention to finite geometries and demonstrate that the F-D + FEM method is one of the most comprehensive approaches presently available to calculate thermal magnetic noise from finite conductors.  To the best of our knowledge, it is the only method that does not impose restrictions on its validity due to some aspects of the experimental setup such as the frequency of measurement, thickness of the conductor, size of the geometry, or suffer from difficulty in practical implementation.

\begin{figure}
    \centering
    \includegraphics[width=0.48\textwidth] {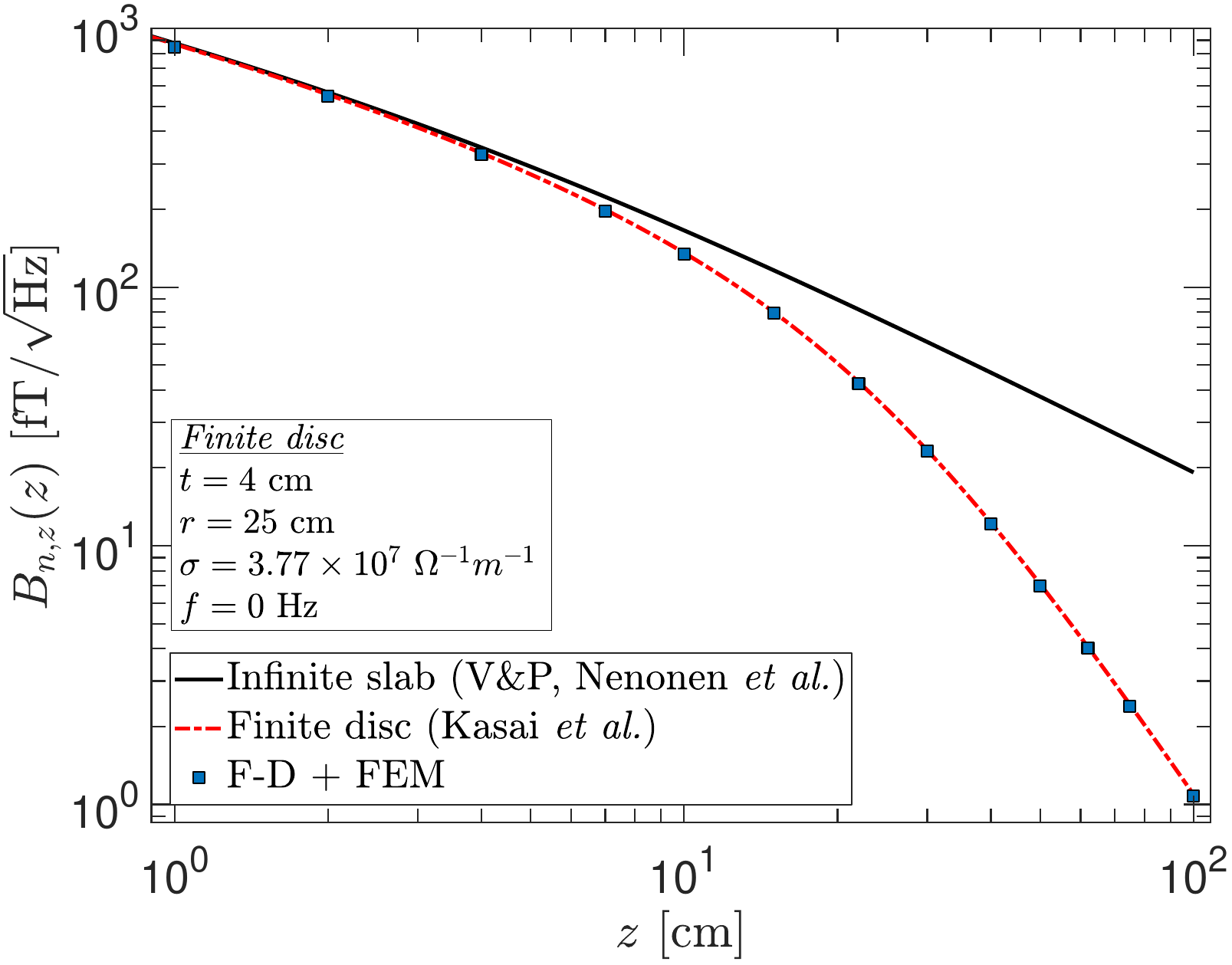}
    \caption{The noise as a function of the distance $z$ along the symmetry axis from a finite "thick" disc of 25~cm radius and 4~cm thickness. The black curve shows the results from approximating the geometry as an infinite slab of the given thickness (Varpula \& Poutanen\cite{Varpula1984}).  The red dashed-dotted curve is the analytic results for a finite disc (Kasai \textit{et al.}\cite{Kasai1993}). The square markers show the results from the F-D + FEM method for this finite geometry.  All calculations are for zero frequency.}
    \label{fig:finite-disc-vs-z}
\end{figure}

We consider, here, a conductor geometry that is frequently encountered in experiments, that of a finite disc or sheet.  Figure~\ref{fig:finite-disc-vs-z} shows the noise calculated for a finite disc of 25~cm radius and 4~cm thickness for observation points at varying distances along the symmetry axis from the conductor surface.  Shown in the figure are the results from three different calculational methods: numerical integration of the Varpula and Pountanen\cite{Varpula1984} expressions for an infinite slab, the analytic formula from Kasai \textit{et al.}\cite{Kasai1993} for a finite disc at zero frequency, and the F-D + FEM method.  The latter has excellent agreement with the analytic results, and as one would expect, the error in the infinite slab approximation to the finite disc becomes significant when the conductor surface-to-observation point distance is comparable to the disc radius.  Although we omit presenting the results, here, a similar comparison of the F-D + FEM method with analytic formulas for the cases of a spherical shell and a cylinder with closed end caps also show agreement.  In Sec.~\ref{sec:application}, we demonstrate instead a practical application of using the F-D + FEM method to compute the noise from a Dewar and magnetically shielded room.

\begin{figure}
    \centering
    \includegraphics[width=0.48\textwidth] {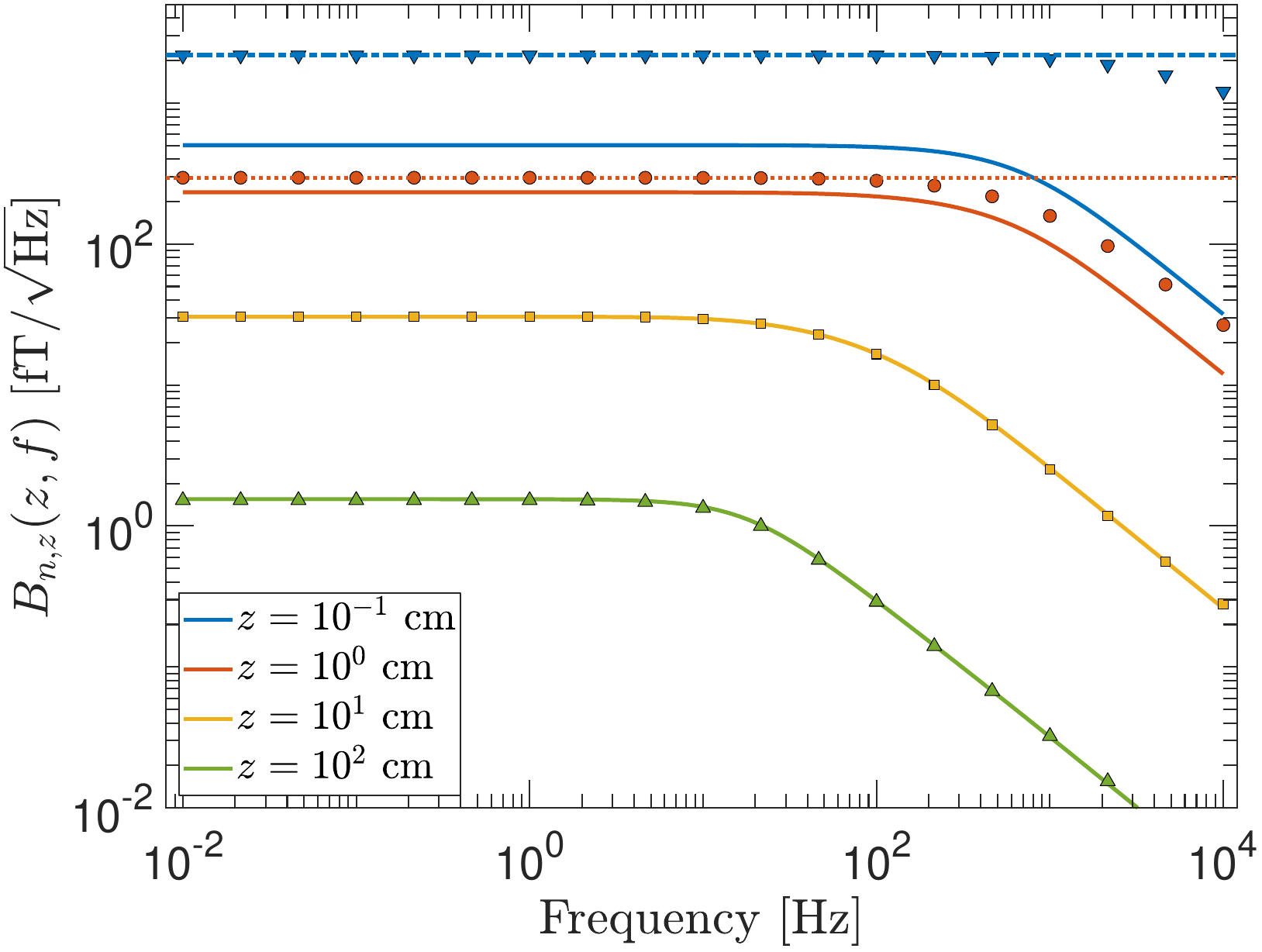}
    \caption{The noise vs frequency at several distances along the symmetry axis from a finite "thin" disc of 100~cm radius and 1~mm thickness. The solid curves are calculated using {\tt bfieldtools}.~\cite{Makinen2020, Zetter2020, Iivanainen2021, Iivanainen2021code} The points show the results from the F-D + FEM method for the given geometry. There is excellent agreement between the two methods for large $z$ compared to the disc thickness, $t$.  But when $z$ is not much greater than the disc thickness (i.e., $z = 0.1$ and 1~cm curves), the {\tt bfieldtools} results begin to deviate from expectations which are indicated by the blue-dot-dashed and red-dotted lines for $z=0.1$~cm and $z=1$~cm, respectively at $f = 0$~Hz (calculated from the analytic formula of Kasai \textit{et al.}\cite{Kasai1993}).  The results from the F-D + FEM method are valid even for small $z/t$ and are consistent with the analytical calculations.}
    \label{fig:finite-disc-vs-freq-z}
\end{figure}

As the above comparison is made at zero frequency, we now show that the F-D + FEM method also provides the correct frequency dependence for arbitrary finite geometries. In this exercise, we benchmark the F-D + FEM method against the method developed by Refs.~\onlinecite{Makinen2020, Zetter2020, Iivanainen2021}.  Their approach models finite conducting objects as thin conductive surfaces on which divergence-free currents give rise to the magnetic noise.  The surface current density is represented in terms of a scalar stream function. Although their method is relatively general and can accommodate arbitrary finite conductor geometries, the principle limitation is that the applicability is restricted to cases for which the observation point is at a distance much larger than the thickness of the conductor.  

When the geometry does, indeed, satisfy the above criterion, the magnetic Johnson noise can be modeled by the very useful open-source Python-based software called {\tt bfieldtools}\cite{Makinen2020, Zetter2020, Iivanainen2021, Iivanainen2021code} developed by the referenced authors.  We use this tool to compute the noise at various $z$ locations from a 1~mm thick disc of 100~cm radius and compare the results with those obtained using the F-D + FEM method.  As shown in Fig.~\ref{fig:finite-disc-vs-freq-z}, the results from these two methods are in agreement when $z \gg t$ (i.e., the $z=10$~cm and $z=100$~cm curves).  However, discrepancies begin to emerge when $z$ is comparable to $t$ ($z =1$~cm), and these increase drastically when $z = t$. Clearly, the two latter cases are outside the regime of applicability of {\tt bfieldtools}, but the purpose of showing these cases is to highlight that the F-D + FEM method continues to remain valid.  This is demonstrated by comparing its results to the blue-dot-dashed and red-dotted horizontal lines in Fig.~\ref{fig:finite-disc-vs-freq-z}.  These curves show the noise calculated at zero frequency for $z = 0.1$ and 1~cm, respectively, using the analytic formula of Kasai \textit{et al.}.\cite{Kasai1993} The results from the F-D + FEM method are consistent with the analytic formula in the zero frequency limit, but as an additional benefit, also provide the general frequency dependence.

\subsection{Volumes}

In addition to computing the noise for a single finite size coil (Sec.~\ref{sec:finite-loop}) and multiple coils arranged as a gradiometer or magnetometer array (Sec.~\ref{sec:gradiometers}), the F-D + FEM method also provides a straightforward procedure to determine the noise from measurements made with coils of finite volume. The two standard geometries frequently encountered in this type of situation are that of a solenoid and sphere.  We consider these two cases as a demonstration of the method for this application.
\begin{figure}
    \centering
    \includegraphics[width=0.48\textwidth] {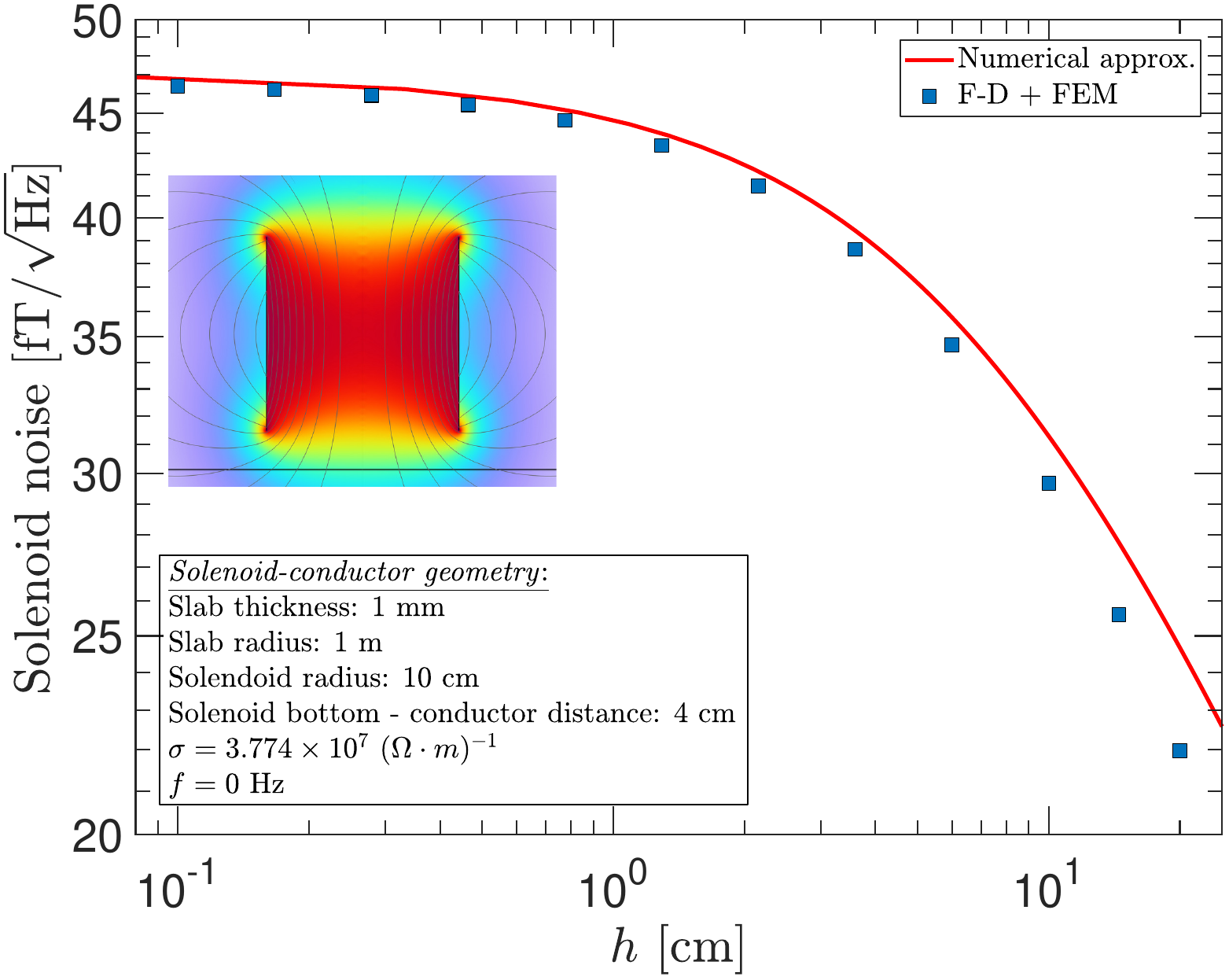}
    \caption{The average noise at zero frequency inside a solenoid as a function of its vertical extent $h$ above a 1~m radius conductor disc with 1~mm thickness.  The red curve shows an approximate numerical estimation of the noise, and the blue-square markers show the results from the F-D + FEM calculation.}
    \label{fig:volume-uniform-solenoid}
\end{figure}

The cylindrical geometry is particularly relevant to many EDM experiments in which a measurement cell, usually of this form, is used to house the particles of interest.  When edge effects are ignored, an approximately uniform field is established inside the cylindrical volume of the solenoid.  The magnitude of the solenoid's magnetic dipole moment is $\vec{p}_{sol} = NI\vec{A}$, where $N$ is the number of turns, $I$ is the current, and $\vec{A}$ is the vector area.  Figure~\ref{fig:volume-uniform-solenoid} shows the noise at zero frequency as a function of the vertical extent of the solenoid $h$ with a fixed radius of 10~cm.  The conductor is 1-mm thick with a radius of 1-m, and the distance from the base of the solenoid to the conductor surface is also fixed.  The red curve is a numerical approximation that considers the solenoid as a stack of thin coils and then uses the results from Fig.~\ref{fig:coil-size-distance} to compute the RMS average. This produces a small overestimation of the noise because not all spatial correlations are accounted for in this approach.  Nevertheless, there is close agreement between these two independent calculations, and they converge in the limit $h \to 0$ as expected.

\begin{figure}
    \centering
    \includegraphics[width=0.48\textwidth] {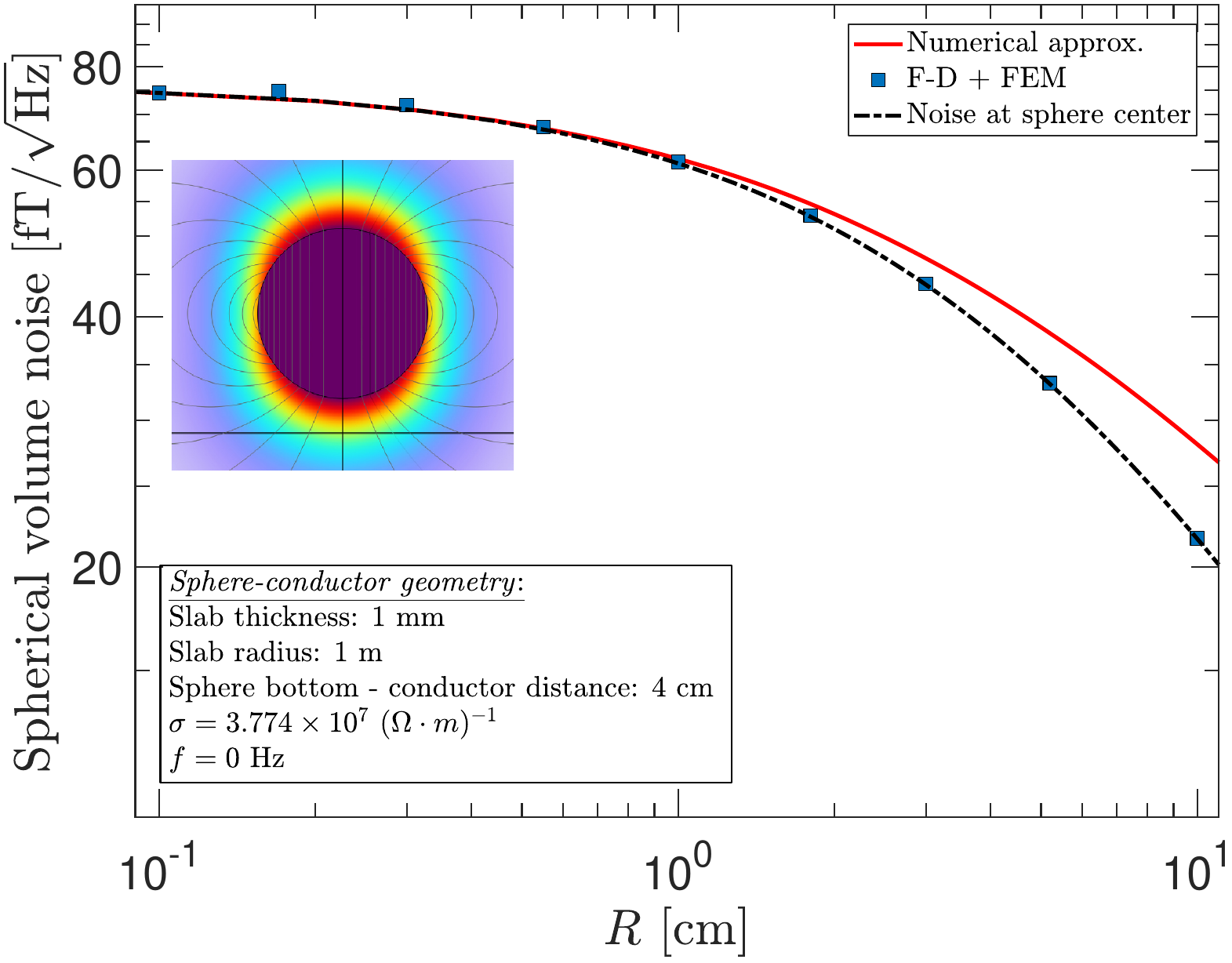}
    \caption{The average noise at zero frequency inside a spherical volume with uniform magnetization as a function of its radius $R$ above a 1~m radius conductor disc with 1~mm thickness.  The red curve shows an approximate numerical estimation of the noise, and the dotted black curve gives the analytical calculation of the noise at the center of the sphere.}
    \label{fig:volume-uniform-sphere}
\end{figure}

The case of a spherical volume is also of practical relevance, in particular, to magnetometers based on a spin-polarized atomic vapor contained in a glass bulb.  This is also a geometry in which an internal uniform field can be created with a current density of a very simple form on the volume surface.  For a sphere of radius $R$ and magnetization $\vec{M} = M_{0}\hat{z}$, the current density is $\vec{J} = M_{0} \sin{\theta} \hat{\phi} = M_{0} (\rho/R) \hat{\phi}$, where $\rho$ is the cylindrical radius, and $\hat{\phi}$ is the azimuthal direction.  The magnetic dipole moment of the sphere is $\vec{p}_{sph} = 4 \pi R^3 \vec{M}/3$, and the field inside the sphere is related to the magnetization by $\vec{B} = 2\mu_{0}\vec{M}/3 $. Figure~\ref{fig:volume-uniform-sphere} shows the average noise inside a spherical volume whose bottom is at a fixed distance from the conductor surface.  The results from the F-D + FEM method are represented by the blue-square markers while the solid-red curve represents a numerical approximation.  The latter is obtained by considering the volume as being composed of many thin discs generated by slicing the sphere along different lines of latitude and then using the results from Fig.~\ref{fig:coil-size-distance} to compute the volume noise.  Similar to the solenoid, this approach produces an overestimation.  However, due to the symmetry of this particular geometry, the average noise can be computed analytically by simply evaluating it at the sphere's center, and this is shown by the black-dotted curve.  Excellent agreement is seen between the analytic results and those of the F-D + FEM method, thus, validating the utility of the latter to volumetric calculations.  But it should be noted that the F-D + FEM method can accommodate more general conditions such as when the geometry is non-symmetric (e.g., when the sphere is offset from the central axis of the conductive disc) or when the frequency of interest is non-zero.

\section{\label{sec:application} Applications}
In this section, as a demonstration of the practical application of the F-D + FEM method, we apply it to calculating the magnetic Johnson noise for two different cases. For both cases, we employ a first-order axial gradiometer consisting of a pair of one-turn circular loops with a diameter of 2.54~cm and a baseline of 5~cm as the coil geometry for  magnetic field detector (Fig.~\ref{fig:gradiometer-geom}(b)). In the first case, the method is used to determine an upper bound on the noise level from the thermal insulation in a fiberglass Dewar. While in the second case, the noise from a magnetically shielded room (MSR) is calculated.

\begin{figure}[htp]
\subfloat[\label{fig:dewar-a}]{
\includegraphics[width=0.74\columnwidth]{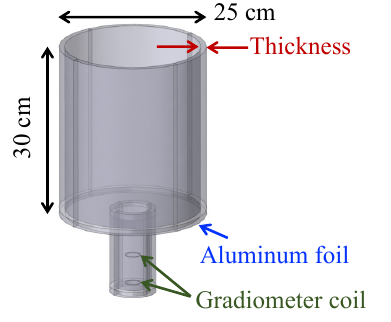}}

\subfloat[\label{fig:dewar-b}]{
\includegraphics[width=0.98\columnwidth]{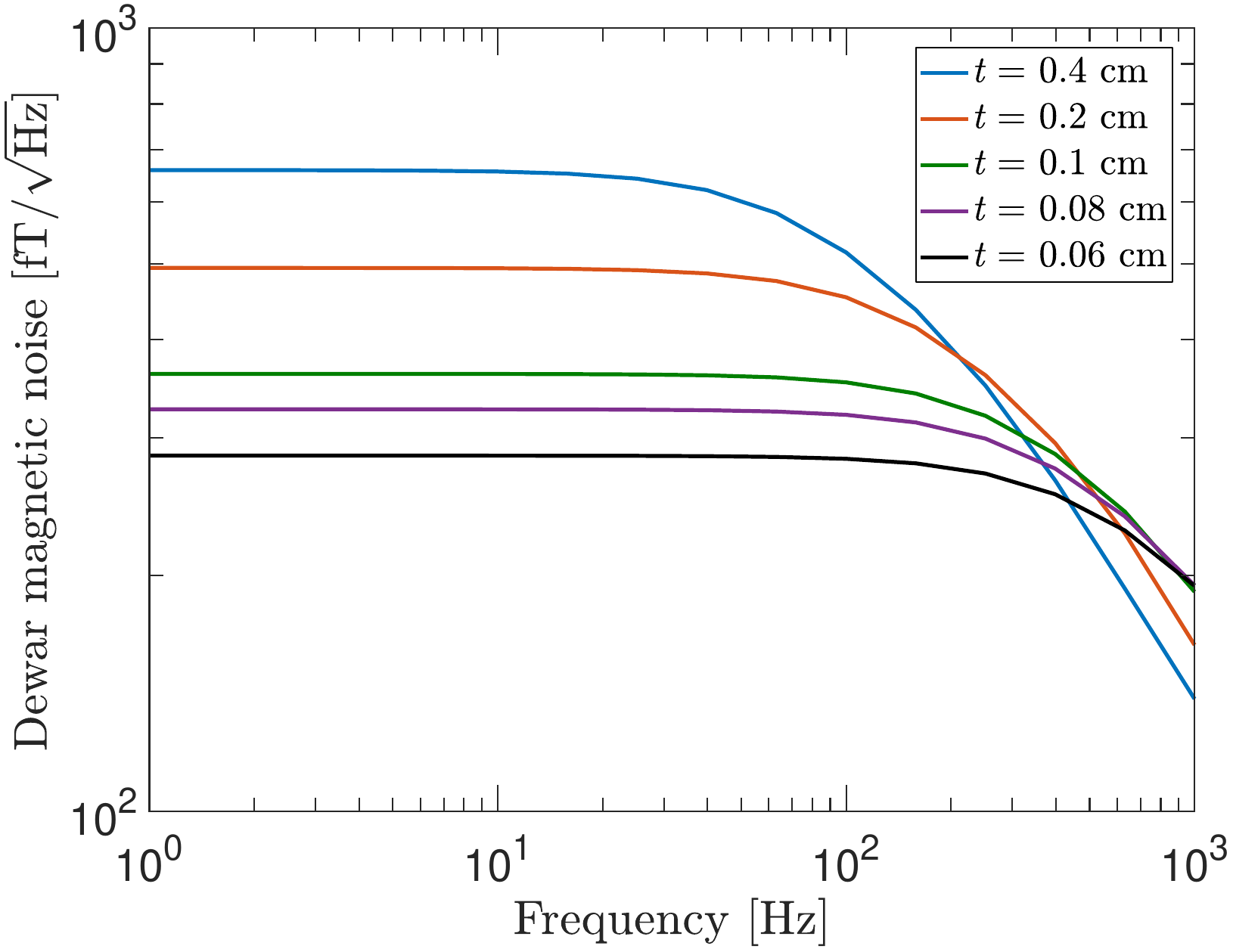}}

\subfloat[\label{fig:dewar-c}]{
\includegraphics[width=0.98\columnwidth]{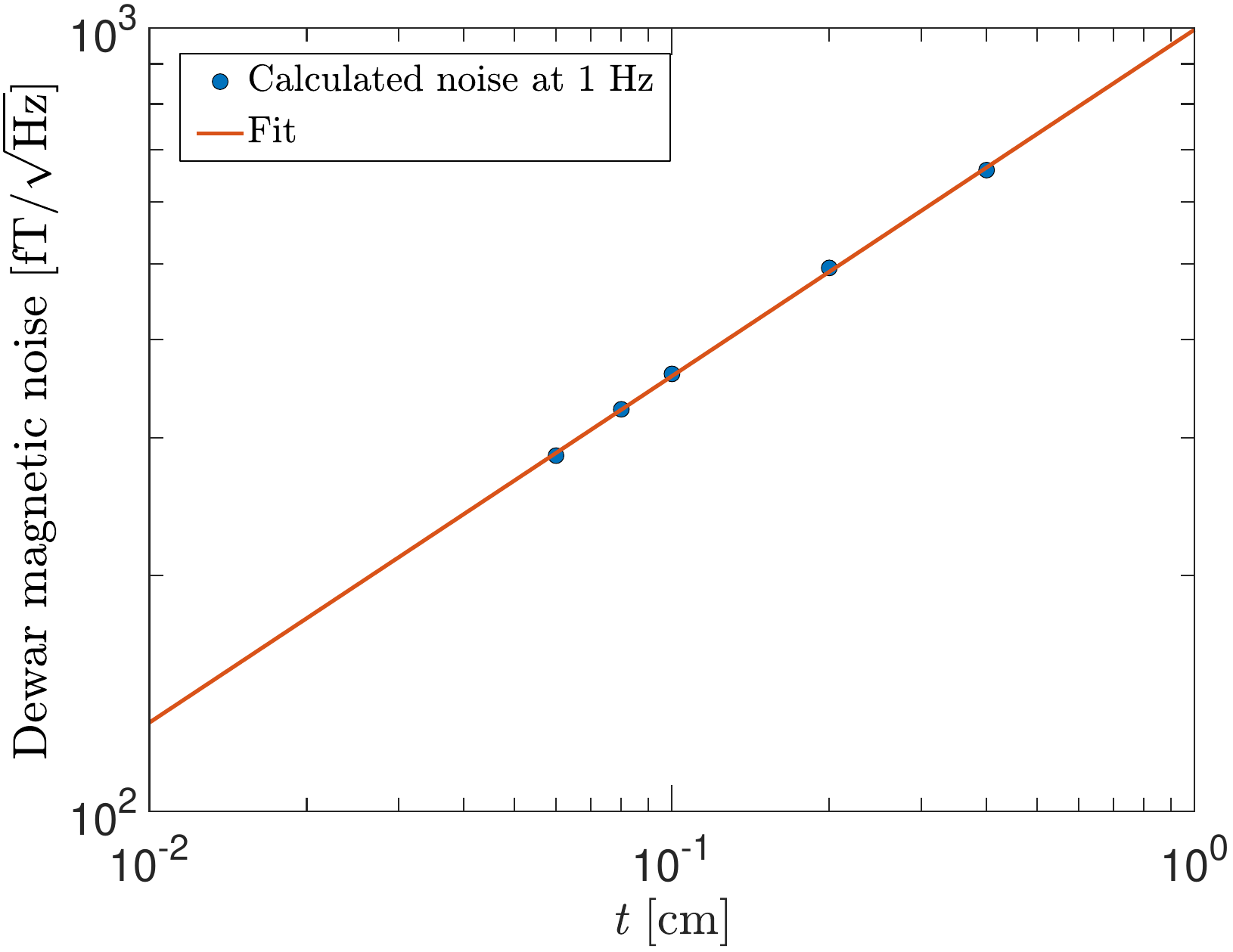}}

\caption{(a) The geometry of aluminum foil surrounding the vacuum space of a Dewar used for the noise calculation. (b) Calculated Dewar magnetic noise as a function of frequency for different aluminum foil thicknesses. (c) Calculated Dewar magnetic noise as a function of the aluminum foil thickness $t$ at 1~Hz. In (c), the solid line indicates a linear fit $\text{log}(y)=a+b\text{log}(x)$, where $a$ and $b$ are free fitting parameters. }
\label{fig:dewar}
\end{figure}

\subsection{Liquid helium fiberglass Dewar}
Many SQUID-based sensitive magnetic field measurements use a liquid helium Dewar made of fiberglass, which typically contains thermal insulation made from multi-layer aluminized mylar film in its vacuum space.~\cite{SETON2005348} The aluminum coating on the mylar film, with a thickness of a few tens of nanometer (nm), dominantly generates the magnetic Johnson noise in this situation. Here, we consider a small-size fiberglass Dewar with the first-order axial gradiometer placed inside. As illustrated in Fig.~\ref{fig:dewar-a}, we use a simple geometry where all of the inner surfaces of the Dewar were lined with thin aluminum foil" and with the gradiometer’s bottom loop separated by 0.5~cm from the nearest aluminum surface. This geometry of thermal insulation gives the largest magnetic noise (worst case scenario), which can be reduced by an improved geometry, e.g., a woven aluminum mylar film that breaks conducting paths via small, self-defined metallization areas.~\cite{SETON2005348} Figure~\ref{fig:dewar-b} shows the calculated Dewar magnetic noise for the gradiometer in the frequency range from 1~Hz to 1~kHz with different thicknesses of aluminum foil. It indicates that the noise reaches a maximum and is attenuated as the thickness becomes comparable to the skin depth (0.25~cm at 1~kHz), as discussed in Fig.~\ref{fig:slab-thickness-frequency}. Based on Fig.~\ref{fig:dewar-c} showing the calculated noise as a function of the aluminum foil thickness $t$ at 1~Hz where the largest noise is generated, one can readily extrapolate the maximum Dewar noise at the thickness of interest with a linear fit of the model $\text{log}(y)=a+b\text{log}(x)$, where $a=(2.998\pm0.006)$, $b=(0.442\pm0.006)$, and y is in fT/$\sqrt{\text{Hz}}$ and $x$ is in cm.  For example, at a practical thickness of 250~nm, the noise is estimated to be $\sim$10~fT/$\sqrt{\text{Hz}}$. With an improved thermal insulation geometry, it has been demonstrated that the maximum Dewar noise level can be suppressed to $\sim$fT/$\sqrt{\text{Hz}}$ by using a woven aluminum mylar film or partly removing the film from the Dewar bottom.\cite{SETON2005348,Fedele_2015}

\subsection{Magnetically shielded room}
Magnetic field measurements are often hindered by environmental magnetic noise.  Therefore, to achieve high measurement sensitivity, the suppression of this noise is necessary and MSRs built from multi-layer high permeability metals (e.g., $\mu$-metal~\cite{mu-metal}) have long been employed towards this goal. For example, in brain neuronal activity imaging, which is accomplished by measuring the magnetic fields produced by neuronal electrical activity, a large MSR with a volume of tens of cubic meters is used.\cite{APL_Kim} The magnetic noise from an MSR has several contributions.  One is due to the electrical conductivity of the layers, typically of order $\sim 10^{6}$ $\Omega^{-1} {\rm m}^{-1}$ for the high-permeability layers and $\sim 10^7$~$\Omega^{-1} {\rm m}^{-1}$ for the non-magnetic layer(s).  For the high-permeability layers, there is an additional noise contribution arising from magnetic domain fluctuations.  But as discussed in Sec.~\ref{sec:high-permeability}, the total noise inside the MSR cannot be determined by calculating the noise due to each layer individually and adding the results in quadrature to obtain the total noise.  The F-D + FEM method, however, is well-suited to handle this type of situation in which the total noise must be calculated simultaneously with all layers present.

\begin{figure}[htp]
\subfloat[\label{fig:MSRa}]{
\includegraphics[width=0.90\columnwidth]{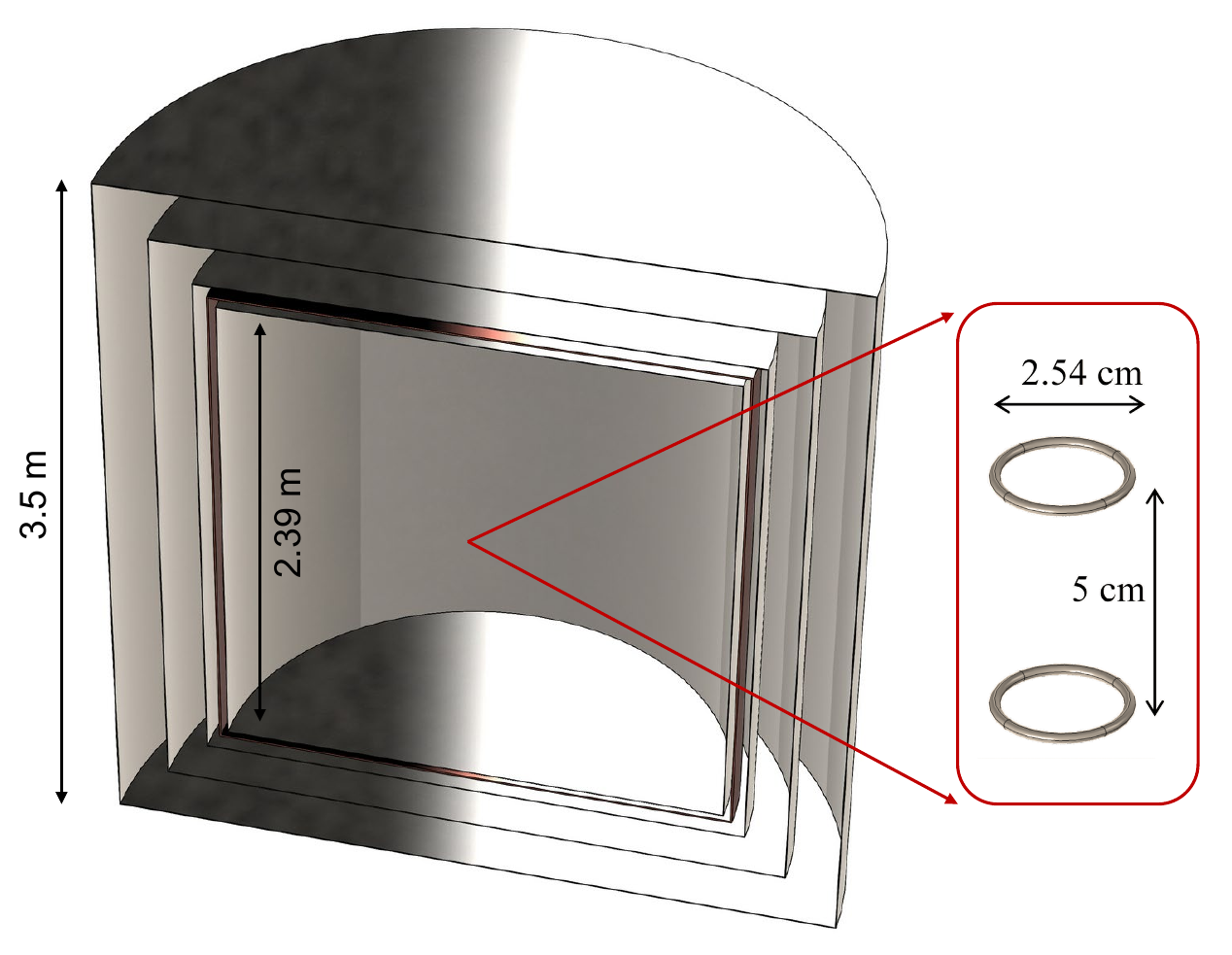}}

\subfloat[\label{fig:MSRb}]{
\includegraphics[width=0.99\columnwidth]{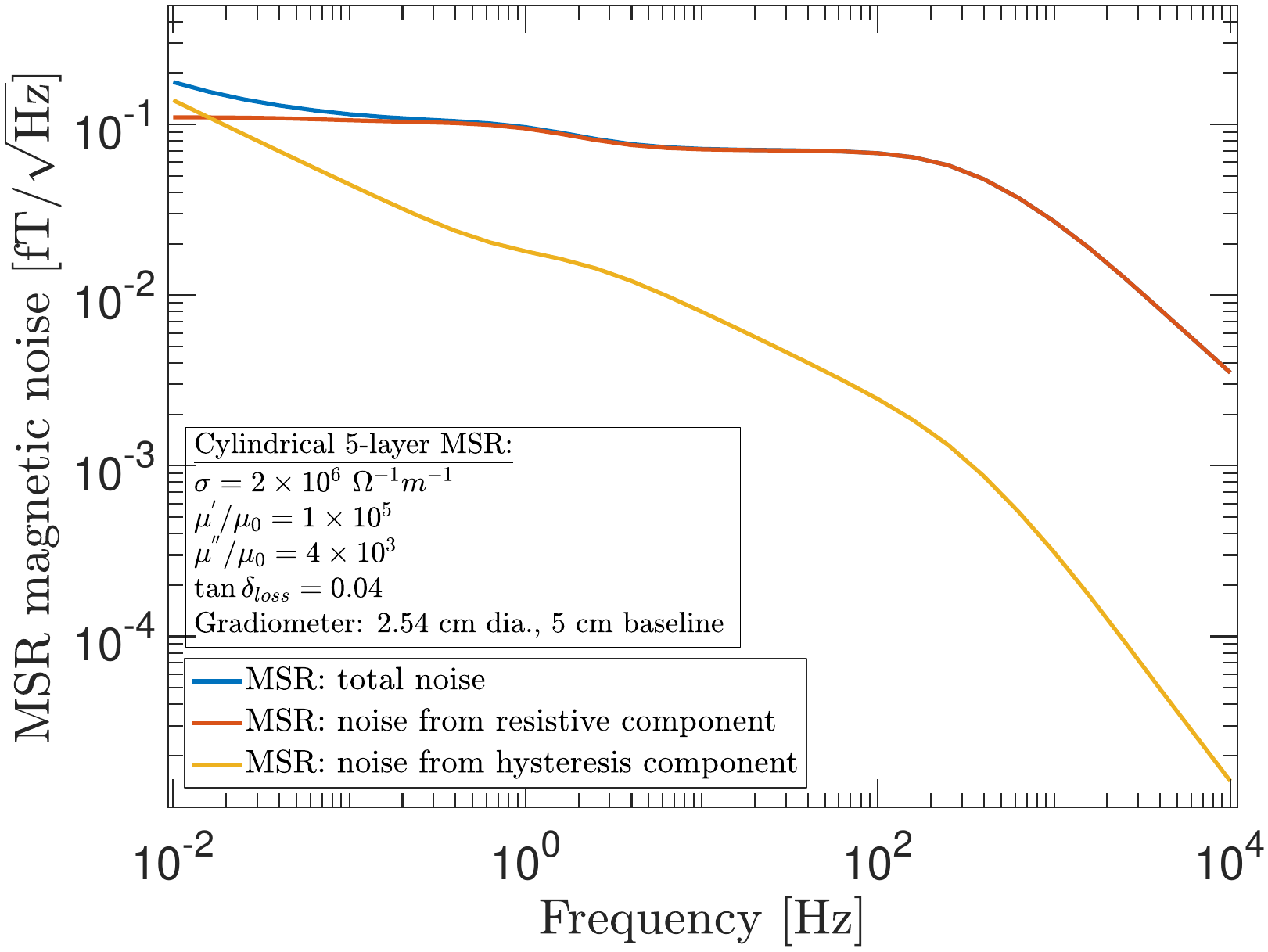}}

\caption{(a) The geometry of a 5-layer MSR with a first-order axial gradiometer placed at its center. (b) The calculated MSR magnetic noise through the gradiometer as a function of frequency.  Shown are the total noise and the contributions from the resistive and hysteresis components.}
\label{fig:MSR}
\end{figure}

The example MSR considered here has a cylindrical geometry with the same height and diameter and is composed of five layers, one copper layer and four layer made from high permeability metal as illustrated in Fig.~\ref{fig:MSRa}. The layers have the following thicknesses starting with the outermost layer: 4~mm, 3~mm, 3~mm, 8~mm (copper), and 2~mm. These values were chosen to be similar to those found in modern multi-layer MSRs, such as the one constructed for the neutron EDM experiment at Los Alamos National Laboratory.~\cite{Ito_2018} The real and imaginary parts of the complex magnetic permeability of the four layers are taken to be $\mu^{'}/\mu_{0} = 10^5$ and $\mu^{''}/\mu_{0} = 4 \times 10^3$, respectively, and the conductivity of the copper layer is taken to be $6\times 10^7$~$\Omega^{-1} {\rm m}^{-1}$.  We calculate the MSR magnetic Johnson noise by locating a first-order axial gradiometer at the center of the room and orientating it to be sensitive to the vertical noise component. Figure~\ref{fig:MSRb} shows the calculated MSR magnetic noise through the gradiometer in the frequency range of $10^{-2}$~Hz to $10^{4}$~Hz, indicating a very small total noise below the 0.2~fT/$\sqrt{\text{Hz}}$ level.  The dominant contributor to the noise is from the innermost layer.  For an MSR-based experiment, this type of calculation would serve as a helpful guide in determining whether the MSR noise could potentially be a limiting factor for the experimental sensitivity.

\section{\label{sec:conclusion} Conclusion}
We have shown in this paper the broad applicability of the method that utilizes a combination of an FEM calculation and the fluctuation-dissipation theorem (the F-D + FEM method) in calculating magnetic Johnson noise. As demonstrated, this approach provides a simple prescription to calculate in a rather straightforward manner the frequency dependence of the magnetic Johnson noise arising from arbitrary conductor geometries with general magnetic field detector configurations. The examples shown included calculations of magnetic Johnson noise from non-magnetic as well as high-permeability conductors.  For the latter class of materials, due to the general property that the total noise cannot be obtained from quadrature summation, FEM methods are likely the best choice to determine the noise for geometries of practical relevance.  Crucially, the accuracy of results obtained through this method relies on a proper account of material properties and power dissipation mechanisms in the regime of interest.

\section{Acknowledgments}
This work was supported by the United States Department of Energy,
Office of Science, Office of Nuclear Physics through Los Alamos National Laboratory under Contract Number 89233218CNA000001 under proposal 2023LANLEED3 and by the Laboratory Directed Research and Development program of Los Alamos National Laboratory under project number LDRD20240078DR.

\section{Data Availability}
Data sharing is not applicable to this article as no new data were created or analyzed in this study.


\section*{References}
\bibliography{johnson_noise}

\end{document}